\documentclass[a4paper, 10pt, twocolumn, nofootinbib,unsortedaddress,superscriptaddress]{revtex4-2} %
\usepackage[utf8]{inputenc}
\usepackage{ragged2e}

\usepackage{natbib}
\usepackage{graphicx} 
\usepackage{setspace}
\usepackage[font=footnotesize,labelfont=bf]{caption}
\usepackage{comment}
\usepackage{float}
\usepackage[caption = false]{subfig}

\usepackage{hyperref} 

\usepackage{enumitem}

\begin{document}
	
	\title{Detect opinion-based groups and reveal polarisation in survey data}
	\author{Alejandro Dinkelberg}
	\email{alejandro.dinkelberg@ul.ie}
	\affiliation{MACSI, Department of Mathematics and Statistics, University of Limerick, Limerick, V94 T9PX, Ireland}
	\affiliation{Centre for Social Issues Research, University of Limerick, Limerick, V94 T9PX, Ireland}

	\author{David JP O'Sullivan}
	\affiliation{MACSI, Department of Mathematics and Statistics, University of Limerick, Limerick, V94 T9PX, Ireland}
	\author{Michael Quayle}
	\affiliation{Centre for Social Issues Research, University of Limerick, Limerick, V94 T9PX, Ireland}
	\affiliation{Department of Psychology, School of Applied Human Sciences, University of KwaZulu-Natal, Pietermaritzburg, South Africa}
	\author{P\'{a}draig MacCarron}
	\affiliation{MACSI, Department of Mathematics and Statistics, University of Limerick, Limerick, V94 T9PX, Ireland}
	\affiliation{Centre for Social Issues Research, University of Limerick, Limerick, V94 T9PX, Ireland}

\begin{abstract}
	Networks, representing attitudinal survey data, expose the structure of opinion-based groups. We make use of these network projections to identify the groups reliably through community detection algorithms and to examine social-identity-based groups. Our goal is to present a method for revealing polarisation and opinion-based in attitudinal surveys. This method can be broken down into the following steps: data preparation, construction of similarity-based networks, algorithmic identification of opinion-based groups, and identification of important items for community structure. We assess the method's performance and possible scope for applying it to empirical data and to a broad range of synthetic data sets. The empirical data application points out possible conclusions (i.e., social-identity polarisation), whereas the synthetic data sets mark out the method's boundaries. Next to an application example on political attitude survey, our results suggest that the method works for various surveys but is also moderated by the efficacy of the community detection algorithms. Concerning the identification of opinion-based groups, we provide a solid method to rank the item's influence on group formation and as a group identifier. We discuss how this network approach for identifying polarisation can classify non-overlapping opinion-based groups even in the absence of extreme opinions.
\end{abstract}

\maketitle

\section{Introduction}
    \label{sec:intro}

Shared opinions are an important feature in the formation of social groups~\cite{kruglanski2006groups}. It has been shown that clusters of opinions become signifiers of group identity~\cite{Bliuc2007}. 
In recent studies, public health opinion-groups have been shown to coalesce around a growing trust/distrust in science, with those having distrust being less compliant with regards to hand-washing and maintaining distance~\cite{maher2020mapping}. This sort of behaviour has major consequences for public health compliance~\cite{vaughan2009effective}.
As a result, it is important to be able to identify such groups accurately, and to reveal if different opinion-based groups are, or will become, polarised on the clusters of topics they share. 

In online communities, such as Facebook groups or subreddit memberships, mutual interests in a subject, or attitudes, are often the primary shared commonality, rather than prior acquaintanceship or geographical proximity. It has been found that in many online communities, users tend to share media aligned with their own values and dismiss alternative views \cite{bakshy2015exposure}. These groups tend to be driven by homophily~\cite{mcpherson2001birds}. 
In this study, we will use this idea of shared attitudes to uncover opinion-based groups from surveys.
In a survey, a participant provides responses on many topics with only a small number of possible response options. These responses are typically on an ordinal scale (e.g., a Likert scale) \cite{DeVellis2003}. Often the scale has a limited range of discrete response options, for example five-point and seven-point scales are commonly employed \cite{groves2011survey}. We use a distance metric, akin to the Manhattan distance, on these scales across all the survey questions, referred to as items, to identify participants with similar opinions. Thus we construct a network of participants linked by shared opinions. This method is discussed in more detail in Ref.~\cite{MacCarron2020}.


There are methods for investigating the social structure and belief networks, see for example~\cite{boutyline2017belief,brandt2019central}. However, in these methods the network is constructed by placing edges between pairs of nodes with a correlation above a certain threshold. This makes the edges difficult to interpret and the choice of threshold is arbitrary. In our approach, we introduce a cut-off when a giant component is formed containing almost all participants. An edge represents shared agreement; the stronger the weight of the edge, the more agreement between these participants.
We use community detection techniques to identify clusters of participants with similar opinions, i.e., opinion-based groups. We compare this to statistical methods, such as hierarchical clustering on the refined survey data and show they give consistent results with each other and, hence, this is a viable method for detecting opinion-based groups and polarisation. 

Surveys can contain hundreds of items, many of which are not expressing attitudes but answering trivial questions leading up to an attitudinal item. We aim to select attitudes, which are closely linked to attitudes of the identified clusters. To do this we apply two feature selection methods to either identify or rank the most relevant items. Ranking the items allows us to reduce the number of items and to highlight influential items.

In this paper we lay out a novel approach on how to remodel attitudinal survey data in order to identify opinion-based groups, applying three different community detection algorithms. We present a new item rank method, which ranks and identifies the items’ importance for an opinion-based group structure. Finally, we demonstrate how to apply this approach to existing data sets.
The paper is laid out as follows, in section~\ref{sec:methods}, we outline the method for forming the networks, identifying the clusters and the feature selection methods for picking the relevant questions. In section~\ref{sec:results}, we show the results and identify the community detection algorithms as robust methods for detecting the opinion-based groups in similarity networks. 
Finally in section~\ref{sec:conclusion}, we discuss the results, give concluding remarks and discuss further research avenues.
    
\section{Methods}
\label{sec:methods}
The detection of opinion-based groups using survey data is conducted in a multi-step procedure. This can be broadly broken down into data restructuring, construction of similarity-based networks~\cite{MacCarron2020}, community detection~\cite{Girvan2002,Karrer2011,Murtagh2011} and item importance. 
 Based on survey data, the method creates links between individuals by constructing a similarity-based network. The emergent structure can reveal opinion-based groups and predict social-group formation \cite{Babeanu2018,Breiger2014}. Once we detect these opinion-based groups, our approach provides a method to rank each item's influence on the opinion-based group structure. 

Survey data often covers multiple contexts with a large number of items. Hence, a subset of items has to be chosen depending on the subject matter of interest. For example, if we focus on political polarisation, then we are interested in identifying politically relevant items which cover attitudes related to party alignment. 
To uncover these attitude connections, we employ a method to project survey data as a similarity network based on the answers of participants. In the resulting network, nodes are participants and weighted links are the similarity scores between participants.

Ref.~\cite{MacCarron2020} shows that the network visualization provides information about groups of individuals that share similar opinions. However, the visualisation and the distinction of groups in the network is highly dependent on layout algorithms, chosen by the user  (in \cite{MacCarron2020}, the Kamada-Kawai layout algorithm~\cite{Kamada1989} was used). A common way in complex networks to partition a graph is the application of community detection algorithms~\cite{Fortunato2010}.
The use of community detection algorithms in our context has the benefit that they do not rely on the visual inspection of the network and that it takes the approach one step further: to reliably 
uncover opinion-based groups.

Over the last two decades a range of different community detection algorithms have evolved (see, for example, \cite{Barabasi2016}). Based on the high complexity of this challenge, there exists no generally applicable algorithm~\cite{Fortunato2010}. The choice of using three distinct algorithms ensures the performance and robustness of the community detection.
Initially we apply the Girvan-Newman algorithm \cite{Girvan2002}, which uses the edge betweenness centrality to minimise the cross-cutting links between communities.
We then run the statistical-driven Hierarchical Clustering algorithm~\cite{Murtagh2011} and finally  the Stochastic Block Model used for community detection~\cite{Karrer2011}. 
%

In the following sections, we explain our approach step by step. 
Although we will later show, using extensive simulations, that our method provides robust results, in order to illustrate the application we first run through a specific example: the American National Election Study (ANES) from 2016~\cite{ANES2017}.
This large data set captures a broad range of general and political attitudes from the American people and includes over 4000 participants and more than 650 items.
We aim to detect opinion-based groups and polarisation in the data set. As an example the ANES data set delivers an ideal candidate to reconfirm polarisation.
Although the ANES data set is not intended to reveal opinion-based groups or polarisation, it captures the particular structure of the American two-party system, which is perceived as bipolar~\cite{Abramowitz2018,Fiorina2008,Iyengar2019}. We take this party alignment as a reference group for community detection and polarisation in this data set.  
For our method, the ANES data set is suitable to investigate polarisation~\cite{Baumann2021}. The first step is to project the survey data as a network.

\subsection{Identifying opinion-based groups from survey data: a score-based linking method}



Attitudinal survey data provides the basis for a network, using the individuals as nodes and their similarity score as links. 
The scales of the items are reformatted into a range between $-1$ and $1$. For instance, a seven-point scale will then be defined as a scale with values of $-1$, $-2/3$, $-1/3$, $0$, $1/3$, $2/3$ and $1$. The scale represents a clear ordinal structure. The reformatting is applied to the whole data set.\\
The similarity measure $S_{ij}$ between the individuals, $i$ and $j$, is the sum of differences between all $n_{f}$ answers to the items $q_n$ (i.e., the Mahattan distance). 
\begin{equation}
S_{ij} = n_{f} - \sum_{f=1}^{n_f} |q_{if}-q_{jf}|.
\end{equation}
The similarity measure ranges between $-n_{f}$ and $n_{f}$. This is to support comprehension, so that an edge with a value of $n_{f}$ displays full agreement between two nodes on all items.
The similarity measure $S$ is at its maximum $n_{f}$, if two individuals have identical responses to all items. Links are drawn, where the similarity exceeds a threshold $\theta$, which is chosen 
when a giant component is formed. Its success criterion is fulfilled if there are enough links in the network to build a giant component, where a path can be drawn to at least 80\% of the individuals. To achieve this, the threshold will successively be lowered until the network matches the success criterion. This means that the procedure includes stepwise links with a lower similarity score.
While the integration of the threshold reduces the number of included individuals, it also reduces the total number of additional links. After these three steps, the data can be shown as a network in order to identify opinion-based groups.
\begin{table}[H]
\centering
\begin{tabular}{l|c|c}
Item                       & Label   & Answer range \\ \hline
Abortion                       & V161232 & 1-4          \\
Race relations                 & V161198 & 1-7          \\
Immigration                    & V161192 & 1-4          \\
Welfare                        & V161209 & 1-3          \\
Homosexuality                  & V161231 & 1-3          \\
Business                       & V161201 & 1-7          \\
Guns                           & V161187 & 1-3          \\
Income                         & V161189 & 1-7         
\end{tabular}
\caption{American National Election Study 2016 - Selected item and their answer range. For a detailed description of the items, see \ref{appendix:ANES_var_descrip}}
\label{tab:ANES2016}
\end{table}%
For the ANES data set, we identified eight items based on a study from \cite{Malka2014} to measure political attitudes. Malka and colleagues' evaluations rely on cultural, economic and self-reported political ideology attitudes.
We then run the data refinement and the network construction on these eight selected items (see Table~\ref{tab:ANES2016}). 
 
\begin{figure*} 
    \centering
    \makebox[0pt]{
    \includegraphics[width=1.1\textwidth]{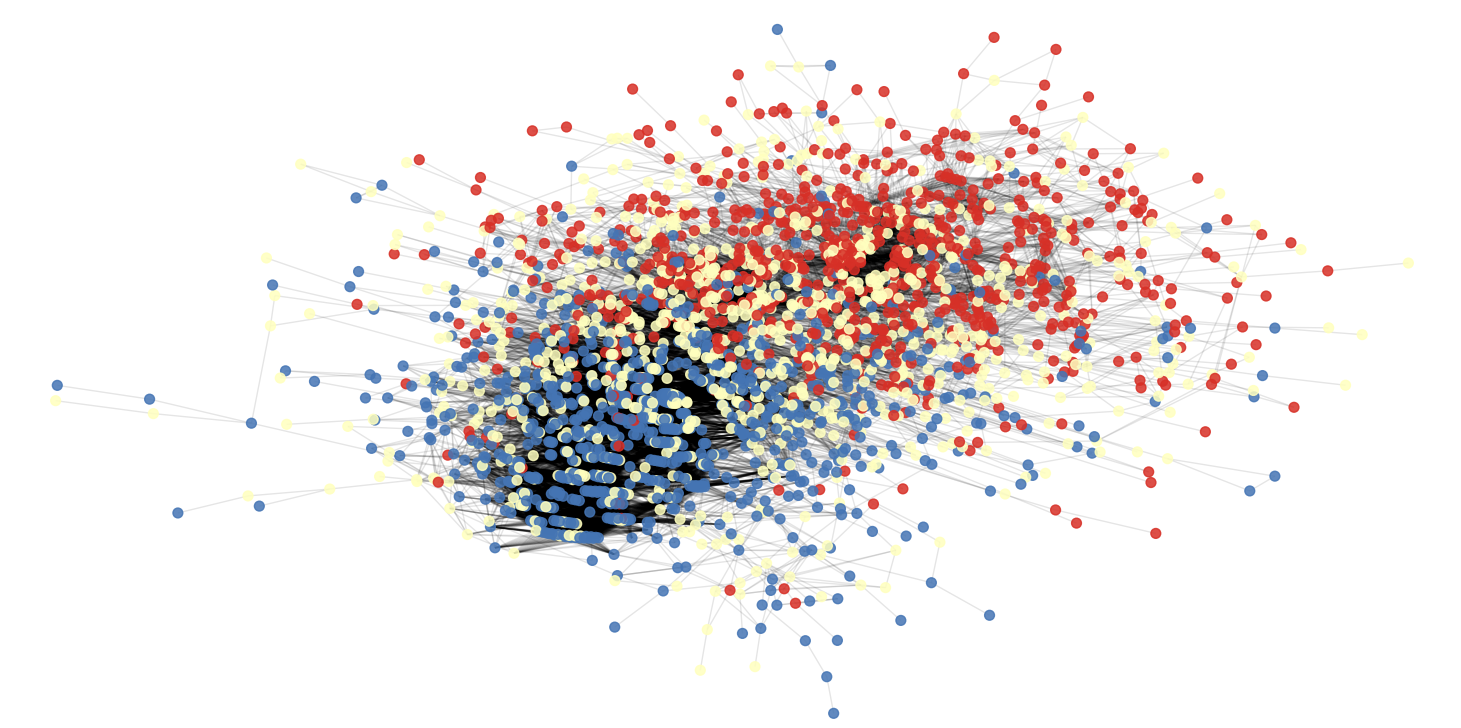}}
    \caption{American National Election Study data 2016, constructed similarity network from the refined data set. The nodes' colour marks the self-identified party affiliation: Republican (red), Democrat (blue) or unknown/independent (yellow).}
    \label{fig:ANES2016}
\end{figure*}

We removed individuals who did not answer all eight items and so our maximum network size here is 3,081 nodes. With a threshold of 7.0, we get 50,143 links between 2,714 individuals, forming a giant component (88.1\% of individuals), where all individuals are connected (see Figure~\ref{fig:ANES2016}).  
%
In our next step, we introduce the community detection for identifying possible opinion-based groups in our network.
\subsection{Detecting opinion-based groups}
Community detection in graphs is an active and ongoing field of research in and of itself, see for example \cite{Fortunato2010,Girvan2002,Javed2018}. Currently there exist a large variety of algorithms to detect group structure from network characteristics~\cite{Fortunato2010,Fortunato2016}. 

In our analysis, we have chosen three different approaches: Girvan-Newman community detection, Hierarchical Clustering and the Stochastic Block Model. The Girvan-Newman algorithm is a network-based method, which is directly applied to our constructed network. In general, Hierarchical Clustering is applied on the refined data set. The Stochastic Block Model is an inference algorithm which detects communities by model fitting. Detailed descriptions of these can be found in the Supplementary Information. 


\subsubsection{Within Sum of Squares}
The Within Sum of Squares (WSS) forms a building block of multiple parts of this analysis, for example, comparing the identified communities of the community detection methods. It is the sum of the squared distance of each individual from their assigned cluster centres.
We can calculate the WSS as follows: 
\begin{equation} 
WSS := \sum\limits_{k=1}^{n_k} \sum\limits_{i\epsilon C_{k}}\sum\limits_{f=1}^{n_f} (q_{if} - \overline{q}_{kf})^2,
\end{equation}
where the number of clusters is $n_k$. $C_k$ is the set of individuals in cluster $k$, $q_{if}$ is individual $i$'s response to item $f$ and $\bar{q}_{kf}$ = $\sum_{i\in C_k} q_{if}/|C_k|$ is the average answer to item $f$ in cluster $k$. 
The goal of our three community detection methods is to reduce WSS substantially while using the least number of communities possible. For two different community assignments, but with the same number of communities, the community with the lower WSS fits better to the data. The preferred partition contains communities, where, on average, the distance between individuals to others in their community is smaller.

With the WSS, we can generate an elbow plot (see \ref{appendix:elbowplot}) for the communities, determined by our community detection methods. An elbow plot displays the WSS versus the number of communities and  gives information about the ideal number of communities in the data~\cite{Yuan2019}.We could use the number of parties that people self-identify as our optimal number of communities. However, the optimal community structure could contain sub-groups within these partisan groups; for instance, we might observe that the community structure is well explained by Republicans or Democrats that are 'centralist', or 'left' and 'right' of the centre, for instance. The elbow plot provides a method for exploring this optimal number of communities. The ``elbow'' in the plot indicates a striking mark for the curve. Successively adding clusters to the data will reduce the total WSS. If the reduction is exceptionally high for an additional cluster, it gives the hint that this might be the ideal number of clusters for the data \cite{Bholowalia2014}. In this way adding more clusters to the data will lead to comparatively small changes in the curve (see in (\ref{appendix:elbowplot}), Figure~\ref{fig:example_elbow}).

\subsubsection{Girvan-Newman algorithm}
The Girvan-Newman algorithm was one of the first community detection algorithms in complex networks \cite{Girvan2002}. This top-down approach divides the network into communities by successively removing links with the highest edge betweenness centrality. Using the edge betweenness centrality, the algorithm intends to identify the community bridging links. It is based on the assumption that links between communities have a higher edge betweenness centrality, caused by their linking ability.

Once the cross-cutting links are identified, they relate two opinion-based groups and mark an attitudinal intersect. The nodes holding these links are positioned at the border of the opinion spaces. Hence, links with a high edge betweenness centrality mark regions where the participants are in the middle of two opinion spaces. The Girvan-Newman algorithm detects and removes cross-cutting links, conceptually dividing the opinion space into smaller, more internally intertwined opinion-based groups.
Also, the edge betweenness centrality is used to measure within community polarisation~\cite{garimella2018quantifying}.

\begin{figure*}
    \centering
    \includegraphics[width=\textwidth]{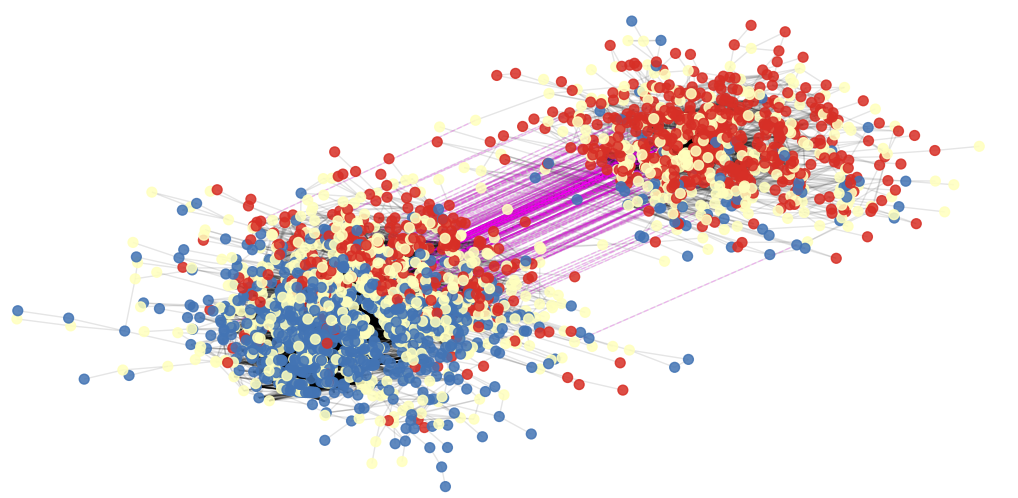}                
    \caption{American National Election Study data 2016, constructed similarity network from the refined data set. With help of the Girvan-Newman algorithm the network is separated into two communities. The purple links are the eliminated links between the communities, and are not part of the network anymore.}
    \label{fig:ANES2016_GN}
\end{figure*}

Our goal is to detect polarisation in the ANES data set from 2016. We run the Girvan-Newman algorithm on the constructed network until it splits into two communities. 
In order to obtain a statement about the overall structure, we re-compute the Girvan-Newman community detection on the biggest community if the first division results in one very small community (less than 5\% the size of the larger community). Applying the Girvan-Newman algorithm to the ANES data set resulted in communities of 41 nodes and 2673 nodes. Running the algorithm again on the larger community, only 190 edges needed to be removed in order to split this community into two communities of 1818 and 855 nodes (see Figure~\ref{fig:ANES2016_GN}). It can be seen that the larger of these components corresponds mostly to democrats (blue nodes) and the smaller to republicans (red nodes).

%
%
\subsubsection{Stochastic Block Model for community detection}
An algorithm for broad range of applications to produce a model to generate networks with community (block) structure is called the Stochastic Block Model~\cite{Holland1983}. 
The model, based on statistical inference, describes the link formation as a process that takes place more often within than between communities. 
The community detection is viewed as a challenge of fitting the Stochastic Block Model to a network in order to reveal a probability-based community structure. Based on this, through an integrated optimisation process a suitable Stochastic Block Model candidate is selected. The flexibility of the Stochastic Block Model means that there exist a variety of approaches for applying and configuring it~\cite{Fortunato2016}. Besides the flexibility, another advantage is the computational complexity in $O(N\ln{N})$~\cite{Fortunato2016}, and therefore the speed of execution is fast compared to the Girvan-Newman algorithm. One drawback of this method, in comparison to the Girvan-Newman algorithm and the Hierarchical Clustering method, is that is is built on stochastic computation. Multiple runs of this method may yield different communities for the same network. It is also not guaranteed that the result is the optimal solution. Nonetheless, Fortunato and Hric \cite{Fortunato2016} assess the Stochastic Block Model as a strong candidate for community detection.

In our approach, we use an algorithm 
in the Python module \textit{graph-tool}~\cite{peixotolibrary2014}. The algorithm provides a degree-corrected or a non-corrected version, which, in our case, will be run according to the minimal description length criterion as described in \cite{Peixoto2017}. 
This function uses an agglomerative heuristic, the Markov Chain Monte Carlo algorithm, for optimisation \cite{Peixoto2014}. The core of the function is a one-dimension minimisation based on the golden section search. More details about the algorithm and its variants can be found in Refs.~\cite{Karrer2011,Peixoto2014,Peixoto2017}.

\subsubsection{Hierarchical Clustering}
 The Hierarchical Clustering method is applied directly to the data set, i.e., without constructing a similarity network. The core of analysis is a distance matrix which contains every distance between the individuals. The distance projects the dissimilarity in their answers over all items. In an iterative process the Hierarchical Clustering merges individuals by aggregating the most similar clusters together, which is decided by a linkage function. In this case, we use a 'group average' linkage function \cite{Everitt2011}. This computes the average of the distance between people in different clusters and aggregates the closest. This leads to the interpretation that the people's clusters who are, on average, close in their opinion are aggregated together initially. Additionally, it is considered to be a intermediate version of the single and compete linkage methods and is relatively robust to outliers. The more common Ward linkage function, which aggregates clusters together that increase the within cluster sum of squares the least, tends to find spherical clusters and is sensitive to outliers \cite{Everitt2011}. However, other linkages functions produce similar clustering results.
The comparison of the three community detection methods arises from the need to choose the ideal number of communities. One approach is to compute a measurement which takes the distances of the answers in each community, the Within Sum of Squares (WSS), into account. The WSS makes it possible to quantify the variability between individuals for a given community assignment. With it, we are able to compare the three methods and, additionally, decide which is the ideal number of communities.
%

\subsection{Selecting relevant items}
Selecting relevant items from large data sets is an important component of our method. Often, to reduce complexity and to include only relevant items, a selection step for the items must be made \textit{a priori}. Therefore, a tool for distinguishing between influential and noisy items would be beneficial to assess the item selection and moreover, rank them in relation to their influence on opinion-based group structure. 

The responses of the survey data constitutes a corresponding vector of opinions for every individual. The differences in their responses form our network structure and opinion-based groups. After the determination of community assignments, we introduce a measurement to locate the relevant items for this particular community structure. Thus, every item is ranked by their meaningfulness. 

The basic concept consists of randomly selecting an item from a data set, shuffling the responses and reallocating them to the individuals. Through this, we break possible correlations to other items and influence on the community assignment if one exists. 
The method is build up as follows:
\begin{enumerate}
    \item The WSS is calculated to obtain a reference value. The calculation is based on the Girvan-Newman, Stochastic Block Model or Hierarchical Clustering community assignments.
    \item At random the method chooses one item and modifies the data set. Consequently, all features are like in the original data set but answers of the selected item are now shuffled.
    \item On the basis of the community assignment, 
    a new WSS is computed. In an additional step, we calculate the ratio of the difference between the old and the new WSS.
    \item To make a reliable statement about the item ranking, the procedures in 2 and 3 is repeated $M$ times per item. In the end, the mean of all WSS-differences is taken to assess each item.
    \item Finally, a value for each item determines the average percentage change of the WSS. Whereas, a higher value means higher influence on the community assignment and values near zero suggest no influence on community assignments.
\end{enumerate}

\begin{figure}
    \centering
    \includegraphics[width=1.1\columnwidth]{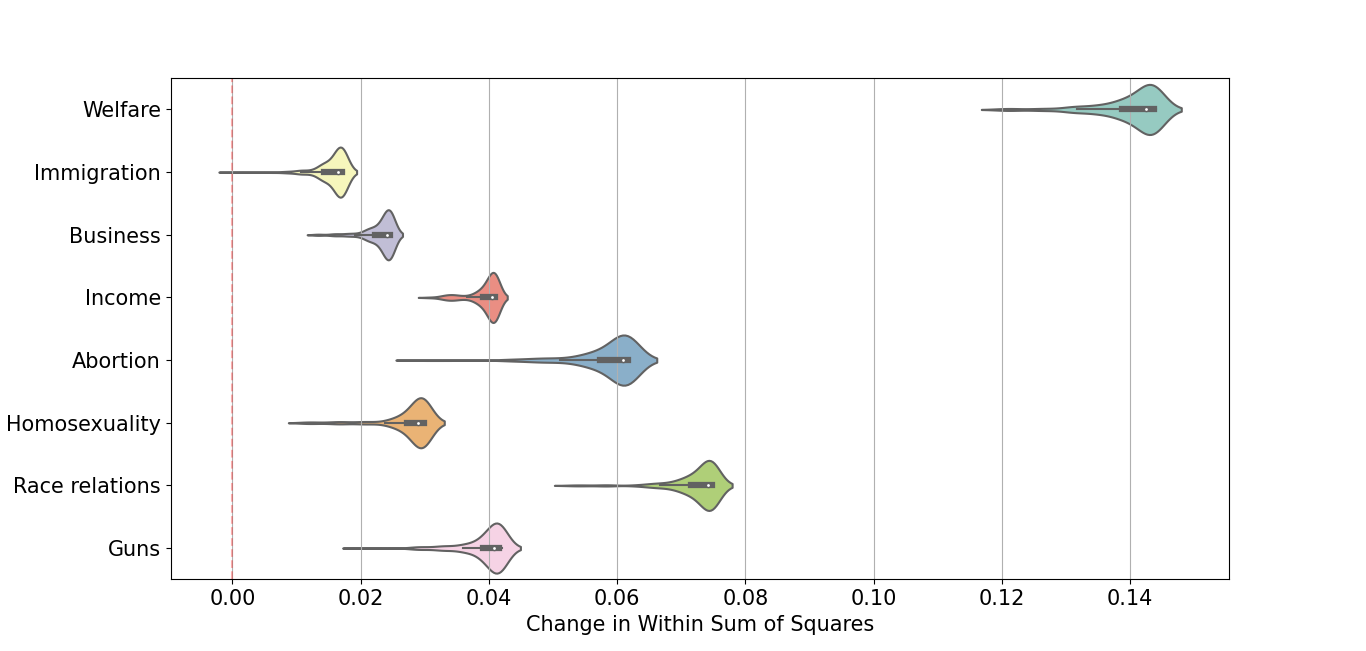}
    \caption{Item selection: Violin plot for the eight items from the ANES data set 2016. It shows the distribution of the average percentage change between the original WSS and the recalculated WSS, in the case of shuffling the items.}
    \label{fig:ANES2016_violinplot}
\end{figure}

The results of the method can be used to produce a violin plot\footnote{It works similar to a common box plot: it marks the median for the WSS-difference for each item, displays the interquartile range and it draws the distribution for WSS-differences using a kernel density estimation.} (see Figure~\ref{fig:ANES2016_violinplot}).\\
Following our example, we computed the item rank method to evaluate the items influence on the community assignments. We ran our method on the eight items from the ANES data set 2016 and simulated it 1000 times, so that, on average, each WSS-distance distribution is based on 125 shuffles of that item. It shows that the item \textit{Welfare} ($V161209$) had the highest and the item \textit{Immigration} ($V161192$) the lowest influence. 
As a comparison for our item rank method we test it against two other methods of feature selection, the \textit{Random Forest} classifier \cite{Breiman2001}, and \textit{Boruta} \cite{Kursa2010}. Random Forests are a substantial modification of the classification trees method that attains near state of the art performance for classification across a wide range of data sets~\cite{Zhang2017,Caruana2006}. 
A Random Forest model is formed from an ensemble of classification trees, where the trees are constructed so they are uncorrelated with each other. A new data point is classified in the model by checking the class that each of the classification trees gives and taking the majority vote of these. The Random Forest model also natively provides item importance measures that can be used to rank the importance of items to the opinion group classification. Please refer to Appendix~\ref{sec:SI_RF} for further details along with useful references to consider when implementing Random Forests.

Boruta is another feature selection method that builds on the Random Forests classifier. It is noted for tackling the 'all-relevant' problem, where, as the name suggests, we seek to find all features that are relevant for the model's ability to classify the opinion-based groups. Several studies have used it successfully as a feature selection tool in a wide range of areas from Fisheries' management~\cite{DiFranco2016} to gene expression~\cite{Kursa2014}. It is a wrapper for the Random Forest algorithm, where it uses a statistical test to identify items that are confirmed to be important, unimportant or undetermined. We are concerned with those items that are deemed to be important to the opinion-based groups under study here. Please refer to Appendix~\ref{sec:SI_rf_boruta} for further details. 

The results of the feature selection for the eight items is shown in Table~\ref{tab:ANES2016_feature_selection}. They are also used in the violin plot (see Figure~\ref{fig:ANES2016_violinplot}) and represent here the average change in the WSS for every item. The second column (Random Forest) shows the values to assess the rank of each item. Evidently, it also ranks \textit{Welfare} and \textit{Race relations} as the two most important items but differs in the rest. The Boruta method defines 7 out of 8 items as important for the community split-up, and validates therefore the selected items for the community detection. Additionally, like the Random Forest method it ranks the item \textit{Gay marriage} as the least important item, whereas the item rank method evaluates \textit{Immigration} as the least important one.

\begin{table}[htb]
\begin{center}
\begin{tabular}{l|c|c|c|c}
{\textbf{Item}} & \textbf{Variable} & \textbf{Item rank} &  \textbf{Random Forest} & \textbf{Boruta} \\ \hline
Welfare     & V161209           & 0.140                             & 0.279                                         & Important                                            \\
Race relations & V161198    & 0.072                             & 0.202                                         & Important                                            \\
Abortion  & V161232        & 0.058                             & 0.134                                         & Important                                            \\
Gun control & V161187      & 0.039                             & 0.062                                         & Important                                            \\
Income  & V161189          & 0.039                             & 0.162                                         & Important                                            \\
Gay marriage & V161231      & 0.028                             & 0.030                                         & Undetermined                                            \\
Business & V161201          & 0.023                             & 0.084                                         & Important                                            \\
Immigration & V161192      & 0.016                             & 0.046                                         & Important                                           
\end{tabular}
\caption{Results for the feature selection by the item rank method, random forest classification and Boruta. The methods were applied on the selected features from the ANES data set 2016, and based on the community detection from the Girvan-Newman algorithm.}
\label{tab:ANES2016_feature_selection}
\end{center}
\end{table}
    
\section{Results}
    \label{sec:results}

In this section, we validate the previously described methods on synthetic data, expand the analysis to new data sets 
and discuss how to apply it to consecutive data sets. 
\subsection{Data sets}
\subsubsection{Synthetic data sets} \label{subsec:syntheticdata}
In this section we compare the three different community detection algorithms and under which circumstances they can be applied. The results will show that the detection of the opinion-based groups is not an artefact of just one community detection method. 
To test our approach, we use simulated survey data, where we specify the ground truth for who belongs to each opinion-based group. Additionally, by building in items that are stronger, weaker or no predictors of group membership, 
we test an item rank method to evaluate the items' influence on the community structure. 
This will provide sound footing for its performance when applying it to real world data sets. 
The application to synthetic data sets will reveals, due to gradual variations of their parameter, the effectiveness of the presented approach. It will also show that the performance of the community detection aligns with our determination of the predefined groups (see \ref{appendix:synthetic_data}). 
%

\begin{figure}
    \centering
    \makebox[0pt]{
    \includegraphics[width=1.2\columnwidth]{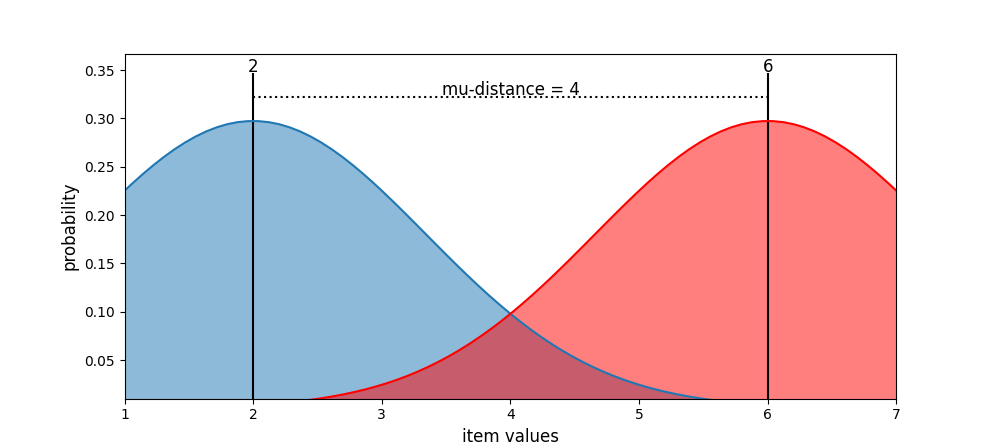}}
    \caption{The \textit{$\mu$-distance} for an item. Group a has mean $\mu_a=2$ and group b has mean $\mu_b=6$. The standard deviation in both cases is $\sigma = 1.4$ }
    \label{fig:mu_distance}
\end{figure}

In the simulated data sets we fix the size of the network, the number of items and the group membership of each individual. An individuals answer to each item is simulated by drawing from a normal distribution with mean $\mu_a$ if they are in group $a$ and $\mu_b$ if they are in group $b$. The standard deviation is the same for the sake of simplicity. The \textit{$\mu$-distance}, $|\mu_a - \mu_b|$,  is a measure of 
the maximal difference the two groups are on an item (see Figure~ \ref{fig:mu_distance}). 

%
\paragraph{Community detection}
The synthetic data sets have 
100 individuals with responses to 7 items on a scale from 1-7. The community structure is an equal division into two groups of size 50. The items are ranked in 4 different categories of information content about the group structure, determined by an increasing standard deviation. 
We consider values of the \textit{$\mu$-distance} from 0.6 to 6.0 (y-axis) and the values of standard deviation from 0.3 to 3.0 (x-axis), both with a step size of 0.3. 

\begin{figure}
\centering
\subfloat[Girvan-Newman]{\includegraphics[width = 8cm]{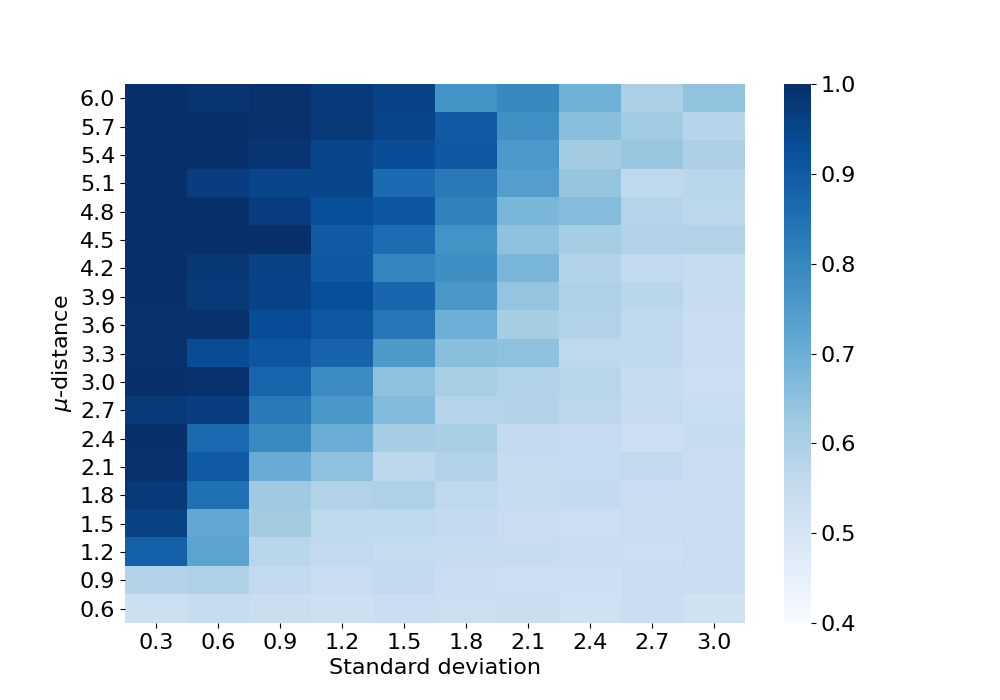}}
\\
\subfloat[Hierarchical Clustering]{\includegraphics[width = 8cm]{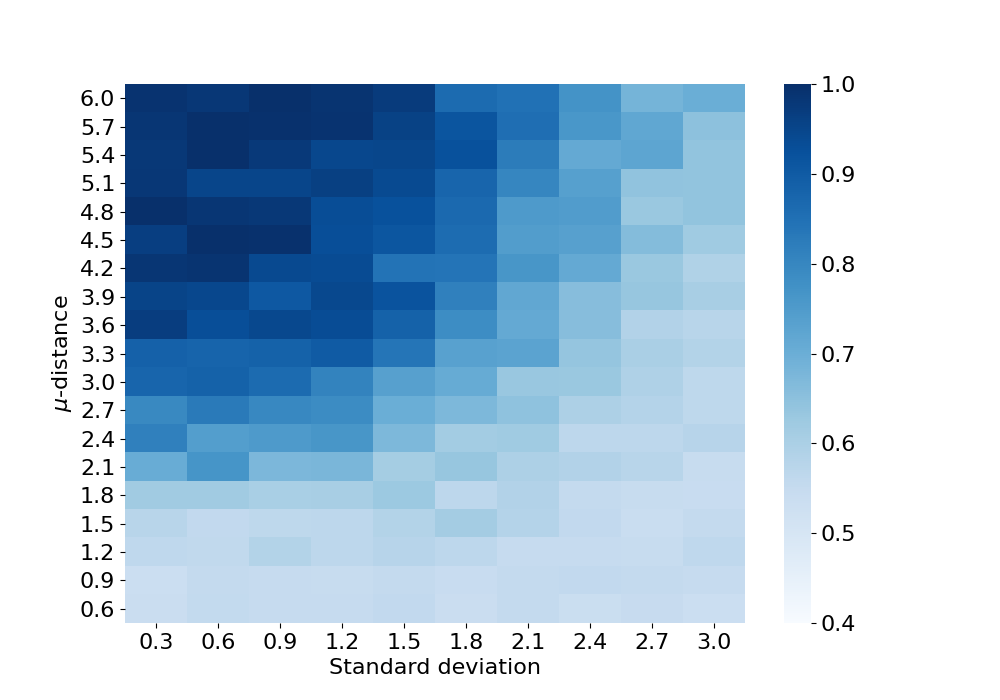}}
\\
\subfloat[Stochastic Block Model]{\includegraphics[width = 8cm]{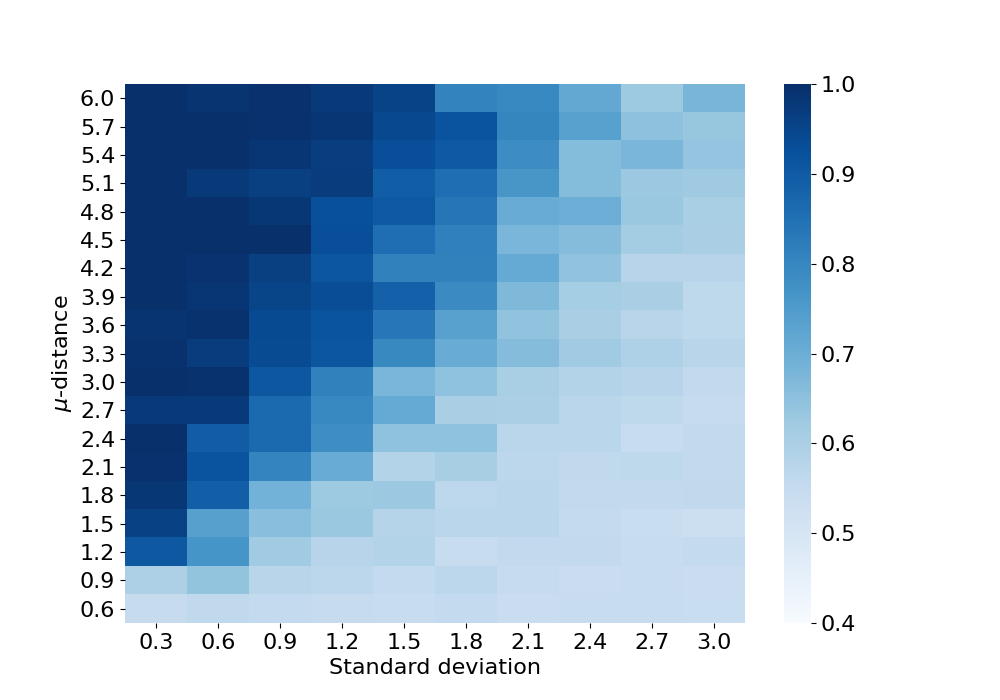}}

\caption{Heatmaps for the mean correct allocation of the community detection algorithms for synthetic data sets, based on 30 runs per parameter constellation.}
\label{fig:heatmaps_synth}
\end{figure}

For every parameter combination we simulate 30 data sets to which we apply the community detection algorithms.
The heatmaps in Figure~\ref{fig:heatmaps_synth} show the percentage of correctly allocated nodes from the network by the community detection algorithms, averaged over the 30 simulations.
%
The heatmaps delineate two regions. The dark blue region is where the community detection works reliably. In this region there is a small overlap in the distributions of item responses for each group due to the high \textit{$\mu$-distance} and the low standard deviation. 
In the light blue region a lower \textit{$\mu$-distance} and a higher \textit{standard deviation} leads to a larger overlap of the responses in the two communities

We also simulated data sets with additional items, which we describe in \ref{appendix:synthetic_data}. If we add additional items there is more information on the group structure, leading to an improved performance of the community detection algorithms (see Figure~\ref{fig:heatmaps_synth_GN}, \ref{fig:heatmaps_synth_HC} and \ref{fig:heatmaps_synth_SBM}). For the shown data set, the difference between the community detection algorithms is minor. A notable difference is only a lack of performance for the Hierarchical Clustering where the \textit{$\mu$-distance} is between 1.2 and 3.3 and the standard deviation is 0.3. The data sets with additional items show similar results. 

\paragraph{Item rank}
To generate the synthetic data sets, items with different levels of information are included. In this way, we provide an order of items, concerning their information about the built-in group structure. The items split up into highly informative, less informative and uniformly distributed noise questions. Likewise the community detection methods, we assess the item rank method by means of synthetic data sets. We test the method’s ability to rank the items correctly by their informational value. In the following, we will see that the item rank method’s performance depends on the ability of the community detection methods to reveal the group structure in the attitude network.  

The bar chart (Figure~\ref{fig:Compare_feature_selection}), representing a cross-section of the heatmaps (see~\ref{appendix:synthetic_data}), shows an equivalent performance of the community detection algorithms. The bar chart captures: a) the proportion of successful community detection in comparison to the ground truth, b) the proportion of correctly detected importance of items, and c) the performance of the \textit{Random Forest} classification algorithm and the $Boruta$ method for feature selection.
By this, it shows what happens in the transition phase, when moving from a dark blue to a light blue region (see~\ref{appendix:synthetic_data},~heatmaps).
\begin{figure*}
\centering
    \makebox[0pt]{
    \includegraphics[width=16cm]{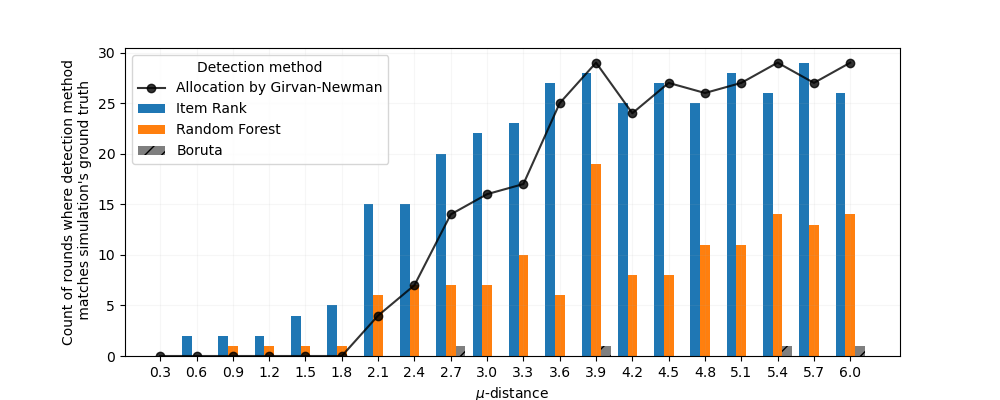}}
    \caption{Results of the correct item ranking, depending on the $\mu$-difference. For each of the item ranking methods (Item Rank, Random Forest, Boruta), each bar shows how often each method detects the correct item ranking for each configuration of the synthetic data set. For every $\mu$-distance we run 30 simulations with 30 different synthetic data sets.
    We added the number that the community detection algorithm (here: Girvan-Newman algorithm) allocated the individuals correctly in relation to the ground truth from the synthetic data set. The graph describes the development of the number of correct allocations for each $\mu$-distance.
}
    \label{fig:Compare_feature_selection}
\end{figure*}
The bar charts show as expected a similar number of correct allocations for the Girvan-Newman algorithm, Hierarchical Clustering and Stochastic Block Model. The number of completely correct ranked items by the item rank method is around 25 out of 30 for the simulations with a maximal~\textit{$\mu$-distance} between 3.3 and 6.0. Below 3.0, the proportion of completely correct rankings of the items decreases. The lack of predictive power of the Random Forest model is of note. It is only able to pick out the correct ranking in a small number of simulations, even when there is a clear community split. The data shows that it performs better in allocating correctly the higher ranked questions but worse in determining the overall ranking. For the same reason, the $Boruta$ method does not determine the rankings accurately. However, it is effective at distinguishing between important and unimportant items.

Applying the approaches on synthetic data sets was beneficial for exploring and comparing the methods under artificial conditions to ensure their robustness in various scenarios. Nevertheless, synthetic data is no substitute for real-world empirical data sets. Furthermore, only by analysing real data sets, inferences can be made and results can be interpreted.

%
%
%
\subsubsection{Wellcome Trust data}
Here, we present a data set from the Wellcome Global Monitor 2018 which has not been previously studied to reveal opinion-based groups. The Wellcome Global Monitor conducted a survey in 2018 in over 140 countries with over 140,000 participants \cite{Wellcome2019}. The survey encompasses public attitudes to science and health. We select attitude-related items from the data set; 10 items deal with trust in organisations, institutions and science, and 3 are attitudes towards vaccines.

We apply our approach to each listed country to detect opinion-based groups and, if applicable, relevant items for community structure. We refine and normalise the data to construct country-specific networks. 
We apply the Girvan-Newman algorithm to detect opinion-based groups, and Hierarchical Clustering to confirm these results. Our approach detects polarisation on health and science attitudes in five countries: Singapore, Venezuela, Cameroon, Congo and Nicaragua (see Table \ref{tab:cutout_wGM2018}, here: \href{https://kurzelinks.de/Analysis-WellcomeGlobalMonitor-2018}{complete Table}).

\begin{table}[!htb]
\centering
\resizebox{\columnwidth}{!}{%
\begin{tabular}{l|c|c|c|c|c|c|c}
Country             & Size & Split-up (GN)   & Links & Removed links & Threshold & Split-up (HC)   & Overlap \\ \hline
Singapore           & 456        & {[}327, 129{]} & 5015              & 12           & 11.5               & {[}326, 130{]} & 0.985    \\ 
Venezuela           & 575        & {[}380, 195{]} & 3757              & 109          & 11                 & {[}452, 123{]} & 0.854   \\ 
Cameroon          & 493        & {[}318, 175{]} & 4767              & 161          & 10                 & {[}401, 92{]}  & 0.542    \\ 
Congo  & 356        & {[}191, 165{]} & 2180              & 47           & 10                 & {[}209, 147{]} & 0.933     \\ 
Nicaragua           & 614        & {[}433, 181{]} & 5466              & 105          & 11                 & {[}423, 191{]} & 0.925    \\
\end{tabular}}
\caption{Results from Wellcome Global monitor in the five countries where we identified polarised opinion-based groups by Girvan-Newman algorithm.}
\label{tab:cutout_wGM2018}
\end{table}%

The most noteworthy result is Singapore. Even though the threshold is very large (11.5), over 5015 links were added to the network, indicating that the individuals have an high consensus within the items. A link between the individuals in the network with a threshold of 11.5 means that in over 90\% their item responses are very close or identical. Next to the high agreement, it was possible to separate the network into two communities by only erasing 12 links. The Hierarchical Clustering provides similar results, with an overlap of over 98.5\% in community allocation. In the other four countries two large opinion-based groups are shown for both community detection methods, with the exception of Cameroon where the overlap of the two method is only about 54\%.

The examination through the item rank method reveals three items as the most important for the communities that we detect via the Girvan-Newman: trust in charity workers, trust in traditional healers and trust in scientists. We showed how to analyse large data sets to uncover the existence of opinion groups. For countries with large opinion-based groups, we are also able to uncover and rank the important items for the community structure.

\subsubsection{Consecutive data sets: ANES 2012 \& 2016}
Polarisation is often seen as an intrasocietal process of moving toward the extremes on political attitudes, e.g., being further away from each others' opinion on a scale. Our method identifies polarisation---even in the absence of extreme opinions---by classifying non-overlapping opinion-based groups.
In the previous section, the ANES data set from 2016 was examined with an item selection based on~\cite{Malka2014}. Here, we investigate the ANES data set from 2012 to demonstrate an approach to consecutive data. Instead of relying on a predetermined selection, we apply the Boruta method to  reveal the important items for our opinion-based groups. To apply Boruta to our data set, we reduced the amount of items from the ANES data set 2016 and 2012 to each 34 items based on relevance (see  here: \href{https://kurzelinks.de/preselection-ANES2012-16}{reduced item list}), selecting  those items obviously related to a personal, political attitude position. Further, we only included participants who self-identified as republicans or democrats as the Boruta method requires group categorisation for the item selection. 
Table \ref{tab:ANES2012_6} shows the items which the Boruta method identifies as important in distinguishing between Democrats and Republicans.
After the normalisation process, the selection of the important items allows us to construct the score-based similarity networks for the ANES data 2012 and 2016. The question is whether the opinion-based clusters are getting more separated, and so easier to detect, or is the opinion-scored network closer together, and therefore it is more difficult to distinguish between communities.

\begin{figure*}
    \centering
    \makebox[0pt]{
    \includegraphics[width=\textwidth]{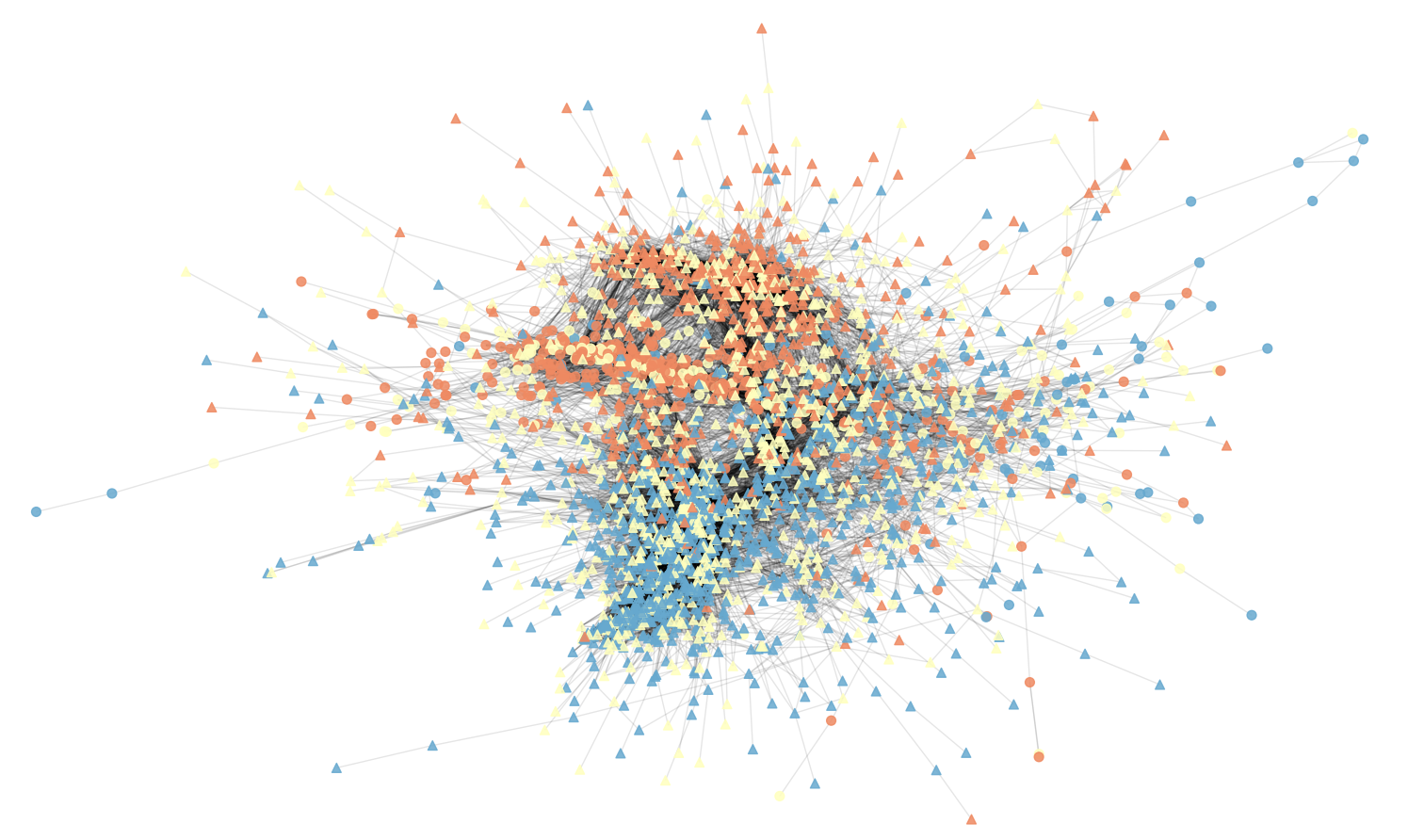}}
    \caption{American National Election Survey data 2012, constructed similarity network from the refined data set, with 2 communities, detected by the Girvan-Newman algorithm. The position and shape of the nodes is used to distinguish between the communities. The colour of the nodes represents their party affiliation: republican (red), democrat (blue) and unknown (yellow).}
    \label{fig:ANES2012}
\end{figure*}

The network for the ANES data set 2012 consists of 2,039 nodes and 31,619 links with a threshold of 8.0 (see Figure~\ref{fig:ANES2012}). The two communities detected by the Girvan-Newman algorithm have 2,493 and 546 nodes. The first community includes 1,004 democrats, 547 republicans and 942 unknown, while the second community consists of 308 republicans, 71 democrats and 167 unknown.

\begin{figure*}
    \centering
    \makebox[0pt]{
    \includegraphics[width=\textwidth]{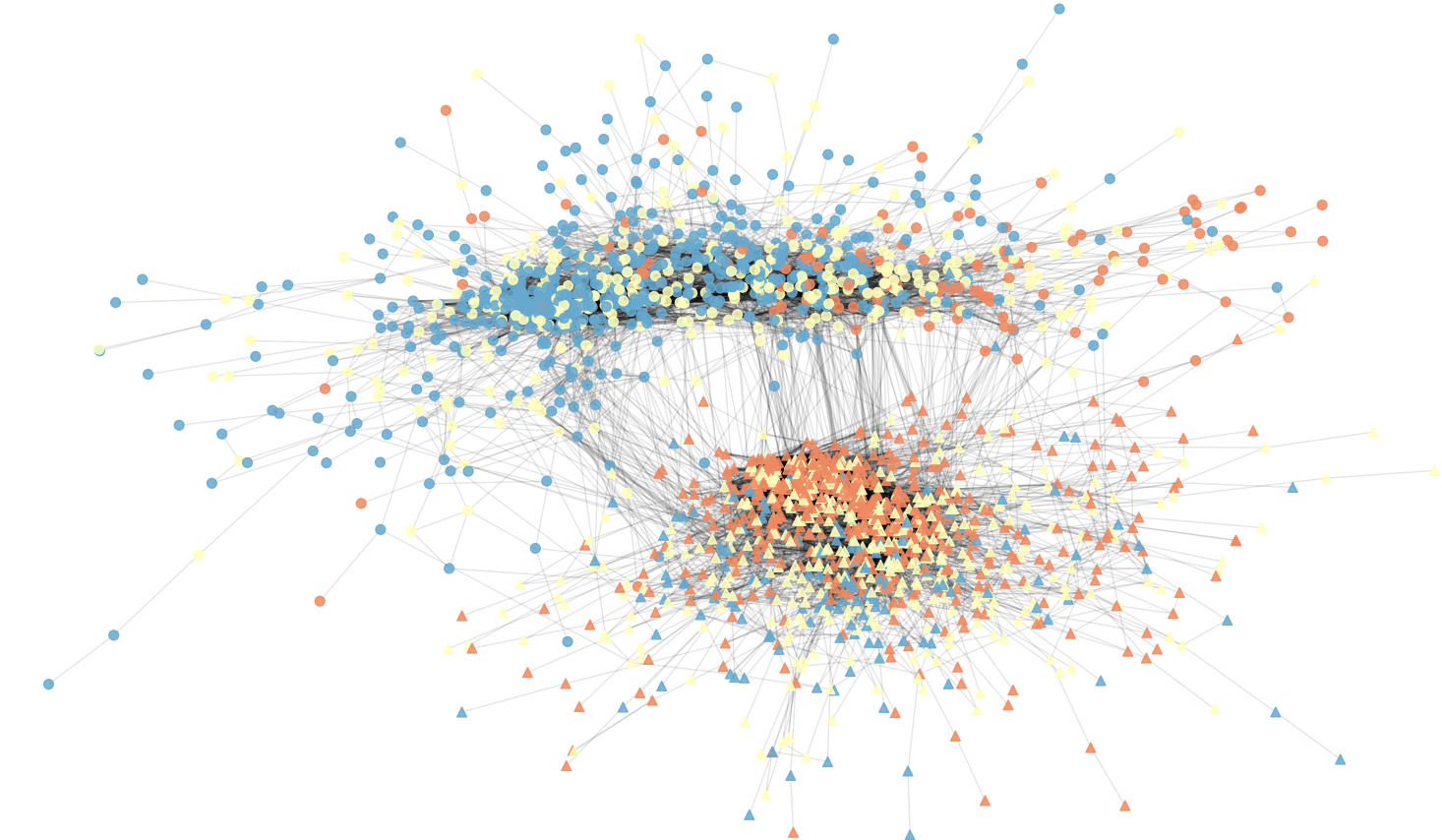}}
    \caption{American National Election Studies data 2016, constructed similarity network from the refined data set, with 2 communities, detected by the Girvan-Newman algorithm. The Boruta method provides the item selection. The network projection uses 10 items from the data set. The position and shape of the nodes is used to distinguish between the communities. The colour of the nodes represents their party affiliation: Republican (red), Democrat (blue) and unknown (yellow).}
    \label{fig:ANES2016_boruta}
\end{figure*}

The network constructed from the ANES data set 2016 includes 2,274 participants and 27,326 links with a threshold of 7.8 (see Figure~\ref{fig:ANES2016_boruta}). Applying the Girvan-Newman algorithm to the network results in a community with 596 republicans, 151 democrats and 424 unknown (total: 1171) and a second community with 612 democrats, 119 republicans and 372 unknown (total: 1103). To reveal the opinion-based groups considerably more links had to be removed in 2012 in comparison to 2016 and the graph had to be re-split several times as it did not fulfil the minimum community size criterion.
\begin{table*}
\centering
\resizebox{\columnwidth}{!}{%
\begin{tabular}{l|c|c|c|c}
Item                       & Label 2012         & Label 2016    & Scale  & Boruta\\ \hline 
Abortion                       & abortpre\_4point   & V161232       & 1-4          & 2012\\ %
Environment-jobs               & envjob\_self       & V161201       & 1-7          & 2012/2016\\ %
Race relations                 & aidblack\_self     & V161198       & 1-7          & 2012/2016 \\ %
Immigration                    & immig\_policy      & V161192       & 1-4          & 2016\\ %
Govt. guaranteed income                         & guarpr\_self       & V161189       & 1-7         & 2012/2016\\ %
Death penalty                  & penalty\_favopp\_x & V161233x      & 1-4        & 2016\\ %
Defence-spending               & defsppr\_self      & V161181       & 1-7           & 2012/2016\\ %
Govt. spending \& services          & spsrvpr\_ssself    & V161178       & 1-7           & 2012/2016\\ %
Medical insurance              & inspre\_self       & V161184       & 1-7           & 2012/2016\\ %
Gay marriage                  & gayrt\_marry       & V161227x       & 1-3/1-6        & 2016\\

\end{tabular}}
\caption{American National Election Studies 2012 and 2016 - Item labels and their answer range. The Boruta method selected the item as important in at least one of the data sets. Only the selected item \textit{Birthright Citizenship} from 2016 is not mentioned due to the fact that there was no corresponding item in 2012. The table show the items that are used for the network projection method and later for the community detection.}
\label{tab:ANES2012_6}
\end{table*}

The application of our opinion-based group detection leads to the conclusion that, based on the ten important items, the ANES data is getting more polarised over time (from 2012 to 2016). The ANES data set from 2016 can be split up by erasing less    cross-cutting links than 2012, and the groups are visually easier to distinguish.

The results show what is already observed: survey participants become increasingly polarised along  party lines on several key opinions~\cite{Iyengar2019} in the ANES data set from 2012 and 2016. The communities are formed around the party affiliation, i.e. each community includes a majority of either Republicans or Democrats. Nevertheless, some participants are more aligned in their attitudes with the other group, contrary to their self-reported partisanship.



\section{Conclusions}\label{sec:conclusion}
In this article, we created a network of individuals linked by similar responses from a survey. We used three different clustering algorithms and show that all are consistent with each other at identifying communities of opinion-based groups on both empirical and simulated data. Further to this, we developed a method to identify the rank and importance of the items in a survey. We compared this to the Random Forest and Boruta method to validate it on simulated survey data. All methods can identify important items, but the method introduced here is more robust at ranking the survey items most important to the identified opinion-based groups. This allowed us to identify which items are most important to the opinion-based group that we found in the ANES and Wellcome Trust data. Additionally, instead of a post-validation, the Boruta method, given a predefined group categorisation, is able to reduce the items of a survey to a subset of group-relevant items.
The exploration of our approach on simulated data also points out limitations for our methods (i.e., opinon-based group detection and item rank). They rely on the performance of the community detection algorithms and, therefore, on the detected communities' meaningfulness.

Being able to identify opinion-based groups is important for understanding a wide range of social issues that can only be solved by the large-scale coordination of opinions (e.g., climate change; public health interventions; vaccination etc.). This is particularly important in understanding online social media interactions, which provide clear affordances for opinion exchange (e.g., via "likes" and "shares"). While identity has been shown to  be central to social opinion processes (e.g., \cite{Doell2021,Gollwitzer2020}), until now it has been difficult to clearly identify links between bundles of opinions and social identities. 

The value of this approach is demonstrated in Ref.~\cite{maher2020mapping} which shows  opinion-based groups emerging at the start of the COVID crisis, progressively polarising on the dimension of distrust in science; and leading to identity-based differences in compliance with public health guidance. We presented a secondary analysis of Wellcome Trust data, identifying countries like Singapore that are highly divided on trust in charity workers and science. 
Similarly, when we analyse the ANES 2012 and 2016 survey data, we identify items in the US that Democrats and Republicans are becoming increasingly polarised on. This phenomenon is widely observed in political and social sciences (see e.g., \cite{Iyengar2019}). 

While we identify separate opinion-based groups here, we do not quantify the level of polarisation, which we aim to address in the future. Network measures to quantify polarisation exist, including using edge betweenness \cite{garimella2018quantifying}. However, these methods all rely on identifying hubs to detect polarised groups. As we construct similarity-based networks, which are dense and weighted, our topology is different. We tend not to have hubs as every node in an opinion-based group will be directly or indirectly linked to every other node in that group
In order to bring methods like this to bear we will need to modify them.

This method for detecting polarisation in opinion-based groups paves the way to investigate the co-constitutive relationship between attitudes and social identity and related phenomena using a network approach. 
\section*{Acknowledgments}
This project has received funding from the European Research Council (ERC) under the European Union's Horizon 2020 research and innovation programme (grant agreement No. 802421). The authors would like to thank Susan C. Fennell for fruitful conversations and feedback. 

\newpage
\bibliographystyle{ws-acs}
\bibliography{bibliography}

\begin{thebibliography}{10}
\providecommand{\urlprefix}{}
\expandafter\ifx\csname urlstyle\endcsname\relax
  \providecommand{\doi}[1]{doi:\discretionary{}{}{}#1}\else
  \providecommand{\doi}{doi:\discretionary{}{}{}\begingroup
  \urlstyle{rm}\Url}\fi

\bibitem{Genuer2010}
{Variable selection using random forests}, \emph{Pattern Recognition Letters}
  \textbf{31} (2010) 2225--2236.

\bibitem{Zhang2017}
{An up-to-date comparison of state-of-the-art classification algorithms},
  \emph{Expert Systems with Applications} \textbf{82} (2017) 128--150.

\bibitem{Abramowitz2018}
Abramowitz, A. and McCoy, J., United states: Racial resentment, negative
  partisanship, and polarization in trump's america, \emph{The {ANNALS} of the
  American Academy of Political and Social Science} \textbf{681} (2018)
  137--156.

\bibitem{ANES2017}
{American National Election Studies (ANES)}, {University Of Michigan}, and
  {Stanford University}, Anes 2016 time series study (2017),
  \doi{10.3886/ICPSR36824.V2}.

\bibitem{Archer2008}
Archer, K.~J. and Kimes, R.~V., {Empirical characterization of random forest
  variable importance measures}, \emph{Computational Statistics and Data
  Analysis} \textbf{52} (2008) 2249--2260.

\bibitem{Babeanu2018}
B{\u{a}}beanu, A.-I., van~de Vis, J., and Garlaschelli, D., Ultrametricity
  increases the predictability of cultural dynamics, \emph{New Journal of
  Physics} \textbf{20} (2018) 103026.

\bibitem{bakshy2015exposure}
Bakshy, E., Messing, S., and Adamic, L.~A., Exposure to ideologically diverse
  news and opinion on facebook, \emph{Science} \textbf{348} (2015) 1130--1132.

\bibitem{Barabasi2016}
Barabasi, A.-L., \emph{Network Science (ONLINE)} (Cambridge University Pr.,
  http://networksciencebook.com/chapter/9

\bibitem{Baumann2021}
Baumann, F., Lorenz-Spreen, P., Sokolov, I.~M., and Starnini, M., Emergence of
  polarized ideological opinions in multidimensional topic spaces,
  \emph{Physical Review X} \textbf{11} (2021).

\bibitem{Bholowalia2014}
Bholowalia, P. and Kumar, A., Ebk-means: A clustering technique based on elbow
  method and k-means in wsn, \emph{International Journal of Computer
  Applications} \textbf{105} (2014) 17--24.

\bibitem{Bliuc2007}
Bliuc, A.-M., McGarty, C., Reynolds, K., and Muntele, D., Opinion-based group
  membership as a predictor of commitment to political action, \emph{European
  Journal of Social Psychology} \textbf{37} (2007) 19--32.

\bibitem{boutyline2017belief}
Boutyline, A. and Vaisey, S., Belief network analysis: A relational approach to
  understanding the structure of attitudes, \emph{American journal of
  sociology} \textbf{122} (2017) 1371--1447.

\bibitem{brandt2019central}
Brandt, M.~J., Sibley, C.~G., and Osborne, D., What is central to political
  belief system networks?, \emph{Personality and Social Psychology Bulletin}
  \textbf{45} (2019) 1352--1364.

\bibitem{Breiger2014}
Breiger, R.~L., Schoon, E., Melamed, D., Asal, V., and Rethemeyer, R.~K.,
  Comparative configurational analysis as a two-mode network problem: A study
  of terrorist group engagement in the drug trade, \emph{Social Networks}
  \textbf{36} (2014) 23--39.

\bibitem{Breiman2001}
Breiman, L., Random forests, \emph{Machine Learning} \textbf{45} (2001) 5--32.

\bibitem{Caruana2006}
Caruana, R. and Niculescu-Mizil, A., {An empirical comparison of supervised
  learning algorithms}, in \emph{ACM International Conference Proceeding
  Series}, Vol. 148 (ACM Press, New York, New York, USA, 2006), ISBN
  1595933832, pp. 161--168, \doi{10.1145/1143844.1143865},
  \urlprefix\url{www.cs.cornell.edu
  http://portal.acm.org/citation.cfm?doid=1143844.1143865}.

\bibitem{DeVellis2003}
DeVellis, R., \emph{{Scale development: theory and applications}}, Applied
  social research methods series (Thousand Oaks: Sage Publications, Inc.,
  2003).

\bibitem{DiFranco2016}
{Di Franco}, A., Thiriet, P., {Di Carlo}, G., Dimitriadis, C., Francour, P.,
  Guti{\'{e}}rrez, N.~L., {Jeudy De Grissac}, A., Koutsoubas, D., Milazzo, M.,
  Otero, M. D.~M., Piante, C., Plass-Johnson, J., Sainz-Trapaga, S.,
  Santarossa, L., Tudela, S., and Guidetti, P., {Five key attributes can
  increase marine protected areas performance for small-scale fisheries
  management}, \emph{Scientific Reports} \textbf{6} (2016) 1--9.

\bibitem{Doell2021}
Doell, K.~C., Pärnamets, P., Harris, E.~A., Hackel, L.~M., and Bavel, J.
  J.~V., Understanding the effects of partisan identity on climate change,
  \emph{Current Opinion in Behavioral Sciences} \textbf{42} (2021) 54--59.

\bibitem{Everitt2011}
Everitt, B.~S., Landau, S., Leese, M., and Stahl, D., \emph{{Cluster analysis:
  Fifth edition}} (2011).

\bibitem{Fiorina2008}
Fiorina, M.~P. and Abrams, S.~J., Political polarization in the american
  public, \emph{Annual Review of Political Science} \textbf{11} (2008)
  563--588.

\bibitem{Fortunato2010}
Fortunato, S., Community detection in graphs, \emph{Physics Reports}
  \textbf{486} (2010) 75--174.

\bibitem{Fortunato2016}
Fortunato, S. and Hric, D., Community detection in networks: A user guide,
  \emph{Physics Reports} \textbf{659} (2016) 1--44.

\bibitem{garimella2018quantifying}
Garimella, K., Morales, G. D.~F., Gionis, A., and Mathioudakis, M., Quantifying
  controversy on social media, \emph{ACM Transactions on Social Computing}
  \textbf{1} (2018) 1--27.

\bibitem{Girvan2002}
Girvan, M. and Newman, M. E.~J., Community structure in social and biological
  networks, \emph{Proceedings of the National Academy of Sciences} \textbf{99}
  (2002) 7821--7826.

\bibitem{Gollwitzer2020}
Gollwitzer, A., Martel, C., Brady, W.~J., Pärnamets, P., Freedman, I.~G.,
  Knowles, E.~D., and Bavel, J. J.~V., Partisan differences in physical
  distancing are linked to health outcomes during the {COVID}-19 pandemic,
  \emph{Nature Human Behaviour} \textbf{4} (2020) 1186--1197.

\bibitem{groves2011survey}
Groves, R.~M., Fowler~Jr, F.~J., Couper, M.~P., Lepkowski, J.~M., Singer, E.,
  and Tourangeau, R., \emph{Survey methodology}, Vol. 561 (John Wiley \& Sons,
  2011).

\bibitem{Ho2002}
Ho, T.~K., {A data complexity analysis of comparative advantages of decision
  forest constructors}, \emph{Pattern Analysis and Applications} \textbf{5}
  (2002) 102--112.

\bibitem{Holland1983}
Holland, P.~W., Laskey, K.~B., and Leinhardt, S., Stochastic blockmodels: First
  steps, \emph{Social Networks} \textbf{5} (1983) 109--137.

\bibitem{Iyengar2019}
Iyengar, S., Lelkes, Y., Levendusky, M., Malhotra, N., and Westwood, S.~J., The
  origins and consequences of affective polarization in the united states,
  \emph{Annual Review of Political Science} \textbf{22} (2019) 129--146.

\bibitem{Javed2018}
Javed, M.~A., Younis, M.~S., Latif, S., Qadir, J., and Baig, A., Community
  detection in networks: A multidisciplinary review, \emph{Journal of Network
  and Computer Applications} \textbf{108} (2018) 87--111.

\bibitem{Kamada1989}
Kamada, T. and Kawai, S., An algorithm for drawing general undirected graphs,
  \emph{Information Processing Letters} \textbf{31} (1989) 7--15.

\bibitem{Karrer2011}
Karrer, B. and Newman, M. E.~J., Stochastic blockmodels and community structure
  in networks, \emph{Physical Review E} \textbf{83} (2011).

\bibitem{kruglanski2006groups}
Kruglanski, A.~W., Pierro, A., Mannetti, L., and De~Grada, E., Groups as
  epistemic providers: need for closure and the unfolding of group-centrism.,
  \emph{Psychological review} \textbf{113} (2006) 84.

\bibitem{Kursa2014}
Kursa, M.~B., {Robustness of Random Forest-based gene selection methods},
  \emph{BMC Bioinformatics} \textbf{15} (2014).

\bibitem{Kursa2010}
Kursa, M.~B. and Rudnicki, W.~R., Feature selection with {theBorutaPackage},
  \emph{Journal of Statistical Software} \textbf{36} (2010).

\bibitem{liaw2002classification}
Liaw, A., Wiener, M., and Others, {Classification and regression by
  randomForest}, \emph{R news} \textbf{2} (2002) 18--22.

\bibitem{MacCarron2020}
MacCarron, P., Maher, P.~J., and Quayle, M., Identifying opinion-based groups
  from survey data: a bipartite network approach, \emph{Preprint}  (2020).

\bibitem{maher2020mapping}
Maher, P.~J., MacCarron, P., and Quayle, M., Mapping public health responses
  with attitude networks: the emergence of opinion-based groups in the uk’s
  early covid-19 response phase, \emph{British Journal of Social Psychology}
  \textbf{59} (2020) 641--652.

\bibitem{Malka2014}
Malka, A., Soto, C.~J., Inzlicht, M., and Lelkes, Y., Do needs for security and
  certainty predict cultural and economic conservatism? a cross-national
  analysis., \emph{Journal of Personality and Social Psychology} \textbf{106}
  (2014) 1031--1051.

\bibitem{mcpherson2001birds}
McPherson, M., Smith-Lovin, L., and Cook, J.~M., Birds of a feather: Homophily
  in social networks, \emph{Annual review of sociology} \textbf{27} (2001)
  415--444.

\bibitem{Murtagh2011}
Murtagh, F. and Contreras, P., Algorithms for hierarchical clustering: an
  overview, \emph{{WIREs} Data Mining and Knowledge Discovery} \textbf{2}
  (2011) 86--97.

\bibitem{Peixoto2014}
Peixoto, T.~P., Efficient monte carlo and greedy heuristic for the inference of
  stochastic block models, \emph{Phys. Rev. E} \textbf{89} (2014) 012804.

\bibitem{peixotolibrary2014}
Peixoto, T.~P., The graph-tool python library, \emph{figshare}  (2014).

\bibitem{Peixoto2017}
Peixoto, T.~P., Nonparametric bayesian inference of the microcanonical
  stochastic block model, \emph{Physical Review E} \textbf{95} (2017).

\bibitem{Strobl2007}
Strobl, C., Boulesteix, A.~L., Zeileis, A., and Hothorn, T., {Bias in random
  forest variable importance measures: Illustrations, sources and a solution},
  \emph{BMC Bioinformatics} \textbf{8} (2007).

\bibitem{vaughan2009effective}
Vaughan, E. and Tinker, T., Effective health risk communication about pandemic
  influenza for vulnerable populations, \emph{American journal of public
  health} \textbf{99} (2009) S324--S332.

\bibitem{Wellcome2019}
{Wellcome Trust} and Ltd, T. G.~O., Wellcome global monitor, 2018 (2019),
  \doi{10.5255/UKDA-SN-8466-2}.

\bibitem{Yuan2019}
Yuan, C. and Yang, H., Research on k-value selection method of k-means
  clustering algorithm, \emph{J} \textbf{2} (2019) 226--235.

\end{thebibliography}

\newpage
\section*{Appendices}
\appendix
\label{sec:appendi}
\section{Community detection algorithms} \label{appendix:Comdetect}
\subsection{Girvan-Newman algorithm}
The community detection algorithm by \cite{Girvan2002} is a divisive approach which successively separates the network into communities. Contrary to other algorithms like Louvain (modularity optimisation method) \cite{Barabasi2016}, it is based on edge betweenness. In our case, we make use of the unweighted edge betweenness centrality in which the edge weights have no influence on the community detection. Using this measure to calculate the centrality of links, it intends to identify through it the community bridging links. It is based on the assumption that links between communities have a higher edge betweenness centrality, caused by their linking ability. Given this, a high amount of shortest paths go through the links to connect nodes between the communities.
The Girvan-Newman algorithm is structured as follows \cite{Girvan2002}:
\begin{enumerate}
    \item The edge betweenness centrality ranks each link.
    \item The link with the highest edge betweenness centrality is selected and removed from the graph.
    \item All links which were influenced by the removal are selected and their edge betweenness is recalculated.
    \item Step 2 and 3 are repeated until every link has been removed from the graph. In our case, we repeat the steps until we split the graph into the desired number of components and terminate the algorithm then.
\end{enumerate}
The community detection algorithms are not only assessed due to their ability to select communities, but as well by their computational complexity \cite{Barabasi2016}.
The Girvan-Newman algorithm's bottleneck is the repeated calculation of the edge betweenness centrality for every link in the network. Its algorithmic complexity is $O(m^2n)$, where the input involves $m$, the number of links, and $n$, the amount of nodes.
This shows that the performance time is exponentially increasing in relation to the input. Due to the very input-sensitive behaviour, the computational costs limits the usage of this method to networks with a maximum of a couple of thousand nodes.\\ 

\subsection{Hierarchical Clustering}
 The Hierarchical Clustering method is applied directly to the data set, thus without constructing a  network. 
The core of analysis is a distance matrix which contains every distance between the individuals $i$ and $j$. There are various ways of calculating the distance between individuals. Here, we choose the euclidean distance:
\begin{equation}
    d(i,j) = \sum\limits_{f=1}^{n_f} (q_{if}-q_{jf})^2,
\end{equation}
where $d(i,j)$ is the distance square distance between individuals (nodes) $i$ and $j$; and $n_f$ is the number of items and $q_{if}$ is individuals $i$'s response to item $f$. 
The distance between two individuals demonstrates the similarity in their answers over all their items. The calculation of distance and the distance matrix are the essential components for the application of the Hierarchical Clustering.\\
The agglomerative character defines the starting point of the Hierarchical Clustering, each individual is defined as a single, separated cluster. The number of clusters is correspondingly as high as the number of individuals, $N$. From there, the algorithm works as follows:
\begin{enumerate}
    \item Considering the distance matrix, select the pair of closest clusters (minimal distance).
    \item Merge the clusters together and recalculate the distance matrix with the new cluster. The merging of clusters uses the group average linkage function. 
    \item Step 1 and 2 are repeated until there is only a giant component left that contains all individuals.
\end{enumerate}
%
%

\section{Feature selection methods} \label{appendix:featureselection}
\subsection{Random Forest} \label{sec:SI_RF}

As mentioned in the main text, the Random Forest model is a classification method that has attained near state of the art performance for classification~\cite{Caruana2006,Zhang2017}.
Making it an attractive option for classification not only due to its accuracy but  its in built variable importance measures and extensions like Bourta (which is discussed in the following section). The ethos of Random Forests is to build a series of Classification Trees using randomised data and items and then apply the majority vote of this ensemble of trees to classify data. The process to construct a Random Forest is as follows. For each tree, we draw a bootstrap sample, sampling with replacement. Any individuals that are not used to construct the tree are held as validation data (referred to as an ``out-of-bag'' sample). Using the bootstrapped sample, we grow a Classification Tree. This process is identical to normal Classification Trees except for one notable difference. At each split, we randomly select $p$ items from the total set of predictors. Using these $p$ variables, we then choose the best variable and split-point. The randomisation of the training set helps to avoid over-fitting. The randomisation of the predictors selected at each split ensures the trees are uncorrelated; otherwise powerful predictors would be likely to be selected, resulting in each tree in the ensemble providing the same information~\cite{Ho2002}. We repeat this process until we have grown the desired number $F$ of trees. Thanks to the randomisation of the data and predictors, we do not need to prune any of the trees that make up the forest, as would be the case for Classification Trees. The predicted probability for any class is the class's proportion that each member of the forest voted for.

A valuable side-effect of randomly selected variables for each split in the ensemble of Classification Trees is that it gives us native access to variable importance measures \cite{Breiman2001,liaw2002classification}. When a variable is used in a split, the decrease in the Gini node impurity is recorded. We can then rank variables by the average of all the Gini impurity reductions, allowing us to observe which questions are most important in forming an opinion-based group. Care must be taken when interpreting variable importance scores as shown in~\cite{Strobl2007}. They noted that issue's can occur when many categorical variables are included with a diverse range in the number of levels. Impurity measures can favour those with many levels. This is not an issue for us as the number of levels in the categorical variables from any survey remains relatively small and similar to each other. Also, if multiple continuous predictors are used with varying scales this can also make the importance measures unreliable. It was noted in \cite{Archer2008} that for continuous predictors that, although the most powerful predictor were not always given the highest importance, these variables were consistency ranked among the top on a ranking of variable importance. In~\cite{Genuer2010} it was noted that multicollinarity is not an issue in Random Forest when ranking item importance. For example, if we added highly correlated predictors the importance is not diluted between them, the ranking is rather preserved. Both \cite{Genuer2010} and \cite{liaw2002classification} provide useful information on the choice of hyper-parameters; the former noting that \textit{ntree} increases the stability of the importance measures and \textit{mtry} increases the distance between the improvement measures (and the latter reference providing useful implementation notes for practitioners, such as the number of trees used should grow with the number of predictor variables used). Additionally, we note similar in our simulation of synthetic survey data, where for categorical variables, the most powerful predictors were ranked highest.  Also of note, though we obtain a rank for the importance of estimates we do not know where the cut off for where a item becomes \textit{unimportant} occurs, of even if it does. The Boruta algorithm addresses this in the Sec.~(\ref{sec:SI_rf_boruta}).


\subsection{Boruta} \label{sec:SI_rf_boruta}

As mentioned in Sec.~\ref{sec:results}, Boruta is a feature selection method where we are concerned with teasing out all relevant features that are predictive, in our case, of the opinion groups that we have found. Boruta is a wrapper for the Random Forest model, that builds on the easy access to the variable importance measures. Providing a means of identifying which items (variables) are not predictive for classification. Please refer to \cite{Kursa2010} for a more extensive discussion of the algorithm's implementation but we will provide the broads strokes here. 

An iteration of Boruta is as follows: It begins by adding a copy of each item to the data set, where each of these are shuffled randomly. These randomised items are called \textit{shadow features}. The shadow features hold no correlation with the classification that the random forest is trying to build and will provide the benchmark for when an item can be declared important. Variable importance is calculated for each item (including shadow features). We note the shadow feature with the largest variable importance and compare all items importance to it. If a item has a variable importance larger than it, they are declared a 'hit', if not they are declared a 'miss'. This process is repeated multiple times so we get, for each items, the fraction of times it was at hit, $p_h$. 

To find when a item is important or not, we take this fraction, $p_h$, and perform a two tailed statistical test based on the binomial distribution.\footnote{In fact, thanks to the number of iterations of Boruta we can use the t-test based on population proportions.} By default this significance level is set to 5\%. For an item, if we fail to reject the null hypothesis, then the item's importance is indistinguishable from that of the shadow features. As a result we can not say it is better or worse than the shadow features. If we can reject the null in favour of the alternative hypothesis, then the item's importance is different from that of the shadow features. Interpreting the sign of the test statistic yields weather the item is important or unimportant to the formation of the observed opinion based group. This process provides a method of isolating which of the features are important to the formation of opinion based groups that we wish to study. As Boruta is an extension of the Random Forest variable importance measure, providing a way of finding a cut-off for which opinions are important, it inherits the same caveats for application to data. Thus, the points and references from Sec.~\ref{sec:SI_RF} that deal with the reliability of the Random Forest variable importance measures should be kept in mind when using Bourta (scale of continuous predictors, large different in number of categories between variables, etc.).

\section{Example elbow plot} \label{appendix:elbowplot}

For each of the community detection methods that we have used we wish to isolate 1) the number of communities that we have and 2) the relative performance of each method. To do this we can construct an elbow plot. 

This plots the Within-cluster Sum of Squares (WSS) we obtain for a specified community assignment, which we sweep successively through. 

\begin{figure}[H]          
    \centering
    \makebox[0pt]{
    \includegraphics[width=1\columnwidth]{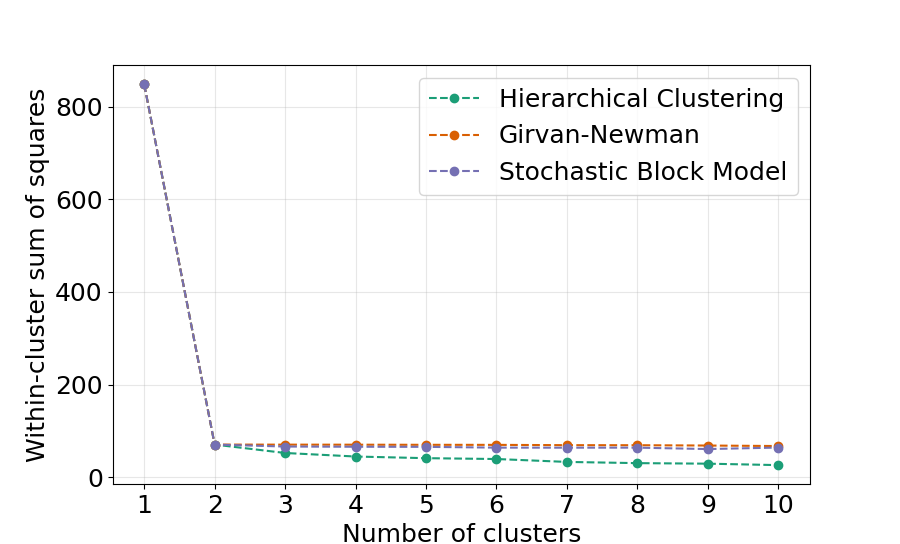}}
    \caption{Example of a successful elbow plot. We generated the plot for a synthetic data set that has two communities by definition. We apply the three community detection algorithms Hierarchical Clustering, Girvan-Newman algorithm and Stochastic Block Model and calculate for each number of community the WSS.}
    \label{fig:example_elbow}
\end{figure}

\begin{figure}[H]          
    \centering
    \makebox[0pt]{
    \includegraphics[width=1.1\columnwidth]{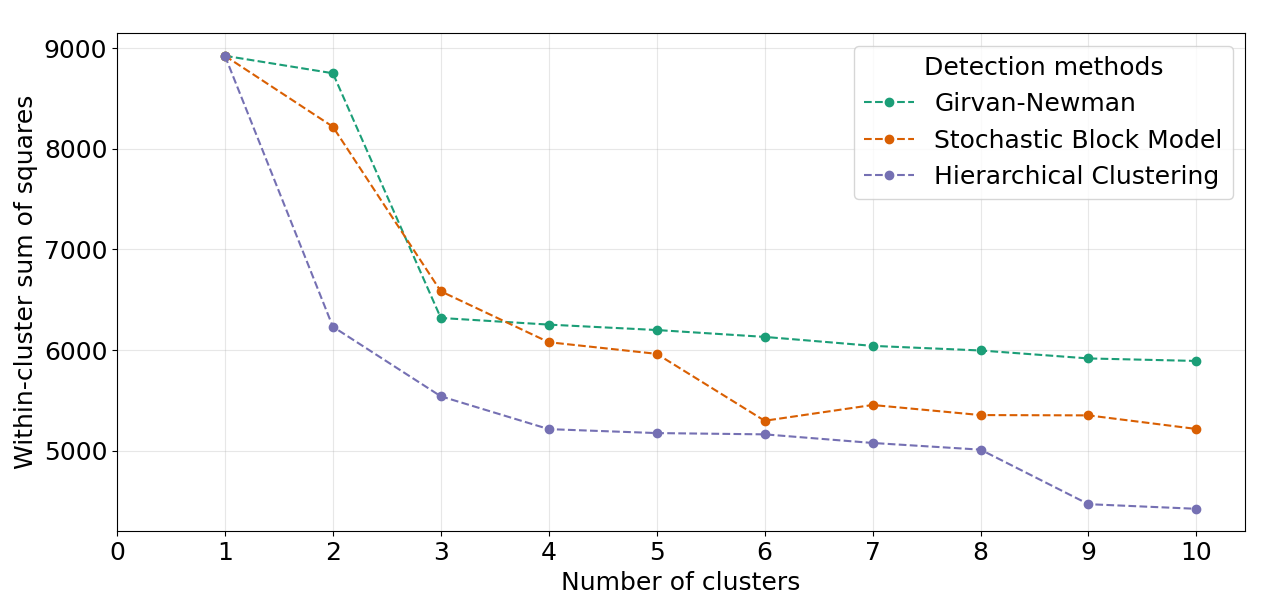}}
    \caption{Elbow plot for the ANES data set 2016 for three community detection algorithms: Hierarchical Clustering, Girvan-Newman algorithm and Stochastic Block Model.}
    \label{fig:tripleelbow}
\end{figure}

If we see an 'elbow' in the plot that point defines the ideal number of communities in the data \cite{Yuan2019}. The elbow defines a point where before it, adding more communities provide large reductions in the WSS; however, after this point, additional clusters only provide marginal reductions in the WSS~\cite{Bholowalia2014}. Beyond the elbow, for a community detection method, additional communities are unlikely to provide increased explanatory power in the community assignment.\\ 
We generate three curves for both synthetic data (Figure~\ref{fig:example_elbow}) and the ANES data (Figure~\ref{fig:tripleelbow}) using the Hierarchical Clustering, Girvan-Newman Algorithm and Stochastic Block Model outlined previously for a range of 1 to 10 communities. Figure~\ref{fig:example_elbow} shows the results for the synthetic data, in which all community detection algorithms behave the same and select similar network communities. The ``elbow'' in the plot yields the correct number of communities as two communities. \\
When we apply this to the 2016 ANES data set using the eight variables discussed in the main text, we find that the picture is similar, if not naturally a little noisier than the synthetic data (see Figure~\ref{fig:tripleelbow}). As you can see, the Hierarchical Clustering provides the lowest WSS than the other methods, and all three confirm that three is the optimal number of clusters. Even though the Hierarchical clustering method offers the lowest number of communities, each method may provide a different but complementary community partition that may be worth exploring. How both the Girvan-Newman and Stochastic Block Model arrive at their community partition leads to a valuable interpretation of these identity-based groups being separated by a small number of tenuous links to the other groups. As such, differences in the beliefs held between these groups might be of interest in-of-themselves. It is also worth noting that using the WSS as a criterion for model selection may be overly harsh on the network-based community detection methods. The Hierarchical Clustering algorithm is built to minimise the distance matrix between communities, so if performing adequately, it should have the lowest WSS.


\section{Description of selected items from the American National Election Study 2016} \label{appendix:ANES_var_descrip}
The selected items are assigned to eight different categories. The ANES 2016 provides a full description of the items (see \href{https://kurzelinks.de/user-guide-anes-2016}{ANES 2016 user guide}). Here, we describe the items and the extracted information from the user guide to allow a quick access and understanding of the used items.
\begin{description}

\item [Abortion (V161232)] The item focuses on the position of the legal regulations of abortion. The participant positions themself and chooses between (1) "By law, abortion should never be permitted";  (2) "By law, only in case of rape, incest, or woman’s life in danger"; (3) "By law, for reasons other than rape, incest, or woman’s life in danger if need established"; (4) By law, abortion as a matter of personal choice".
\item [Race relations (V161198)] The participant is asked to evaluate the government's racial re-distributive policies. The participants can choose from an answer categories which range from \textit{Government should help the black people} to \textit{The black people should help themselves}.
\item [Immigration (V161192)] The item is about the government's policy toward unauthorized immigrants. The participant positions themself and chooses between (1) "Make all unauthorized immigrants felons and send them back to their home country"; (2) "Have a guest worker program in order to work"; (3) "Allow to remain and eventually qualify for U.S. citizenship, if they meet"; and (4) "Allow to remain and eventually qualify for U.S. citizenship without penalties"
\item [Welfare (V161209)] The item's topic is the amount of government spending on welfare programs. The question is whether it should be "increased", "decreased" or "kept the same".
\item [Homosexuality (V161231)] The item addresses the position and legal acceptance of gay marriage and allows the participant to choose between "Gay and lesbian couples
should be allowed to legally marry", "Gay and lesbian couples should be allowed to form civil unions but not legally marry" and "There should be no legal recognition of a gay or lesbian couple’s relationship".
\item [Business (V161201)] The item rates the position of the participants towards the potential cost of jobs through implementing environmental regulations. The answers can range from "Regulate business to protect the environment and create jobs" to "No regulation because it will not work and will cost jobs"
\item [Guns (V161187)] The question is whether the should it make more difficult to to buy a gun. The participant's answers are "increased", "decreased" or "kept the same".
\item [Income (V161189)] The item is to determine the participants' attitude towards the governmental effort to job and income. The item's scale ranges from “Government should see to jobs and standard of living” to “Government should let each person get ahead on own “.
\end{description}

\section{Method to create our synthetic data sets} \label{appendix:synthetic_data}
The idea of generating a synthetic data set is to systematically vary properties of the groups in a dataset and apply our method to this data to access its performance.\\
For this method, we can define the following variables required to create simulated data for testing:
\begin{itemize}
    \item $n\_agents$ = number of individuals in the data set
    \item $n\_items$ = number of questions of the created survey
    \item $scale\_steps$ = size of scale for every question. It will be the same for every n questions.
    \item $mu\_max$ = maximal \textit{$\mu$-difference}, which defines the highest difference available in the questions. The questions can have a smaller \textit{$\mu$-difference}, if their ranking is lower.
    \item $number\_ranks$ = number of differently ranked questions in the data set.
    \item $n\_comp$ = number of predefined  communities in the data
    \item $sd$ = lowest standard deviation for the highest ranked questions
    \item $split\_up$ = percentages to define the size of the community in relation to the overall number of the individuals.
\end{itemize}

The number of questions per ranking depends on the number of questions and on the number of ranks and is then normally distributed around an expected value to allow variation. The importance of the questions is set due to a higher or lower \textit{$\mu$-difference}. The higher the importance of the question, the higher the \textit{$\mu$-difference}. The $mu\_max$ defines the \textit{$\mu$-difference} for the questions with the highest ranking, all the other questions will have a lower \textit{$\mu$-difference} or are noise questions. The method constructs a data set for the number of requested individuals and questions. It is structured like the data used in from the ANES 2016 but without missing data points, and therefore meets our requirements of replicating attitudinal survey data.

\subsection{Simulations based on synthetic data}
\begin{figure*}
\centering
\makebox[0pt]{
\subfloat[]{\includegraphics[width = 9cm]{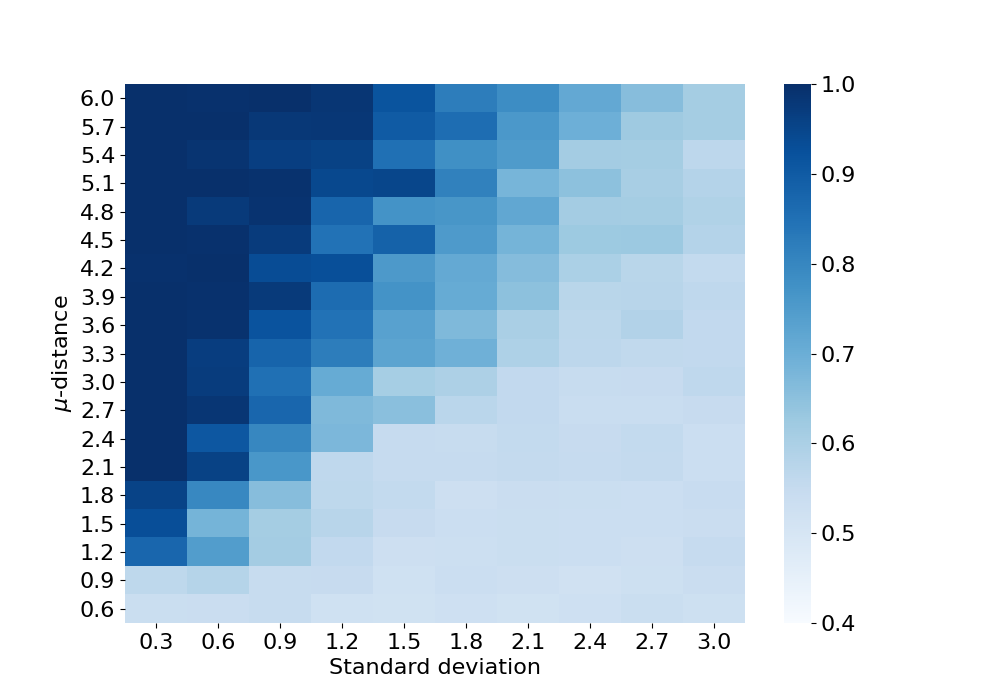}}
\subfloat[]{\includegraphics[width = 9cm]{GN_data_comp_N=100_n_comp=2_total_q=7_noise_q=1_sc=7.png}}}\\
\makebox[0pt]{
\subfloat[]{\includegraphics[width = 9cm]{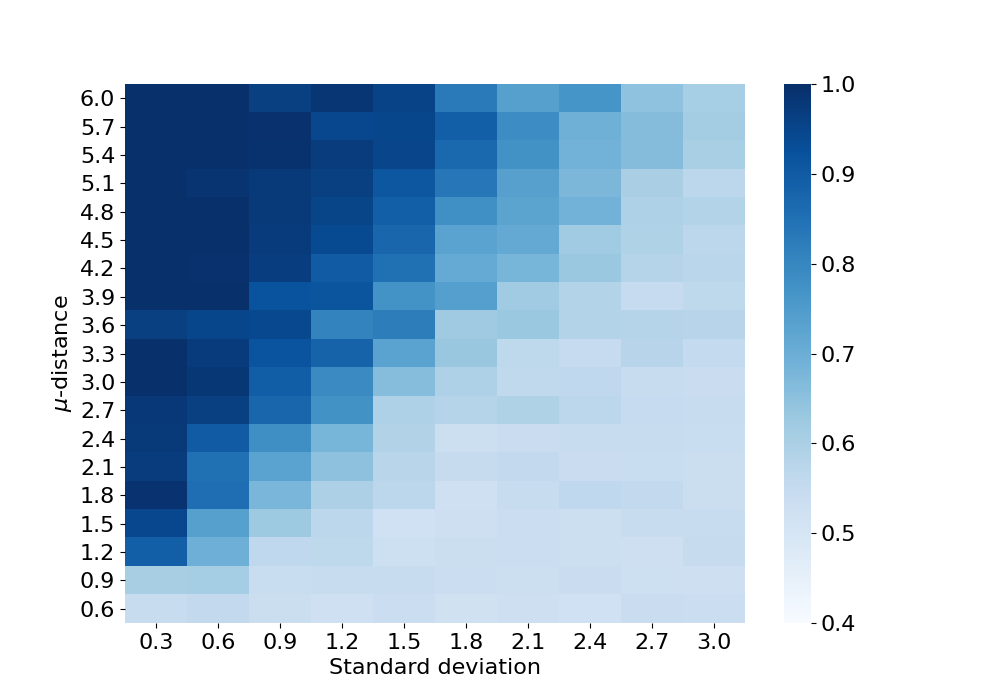}}
\subfloat[]{\includegraphics[width = 9cm]{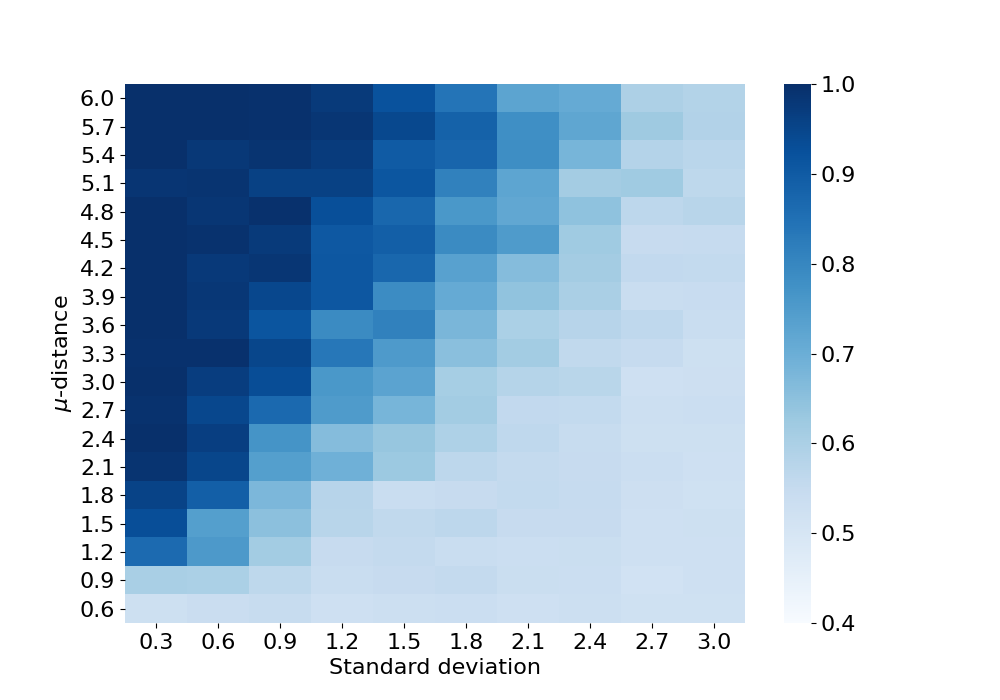}}}

\caption{Heatmaps for the mean correct allocation of Girvan-Newman algorithm for synthetic data sets, based on 30 runs per parameter constellation. The four heatmaps only differ in the number of integrated items: a) 6, b) 7, c) 8, d) 9.}
\label{fig:heatmaps_synth_GN}
\end{figure*}

\begin{figure*}
\centering
\makebox[0pt]{
\subfloat[]{\includegraphics[width = 9cm]{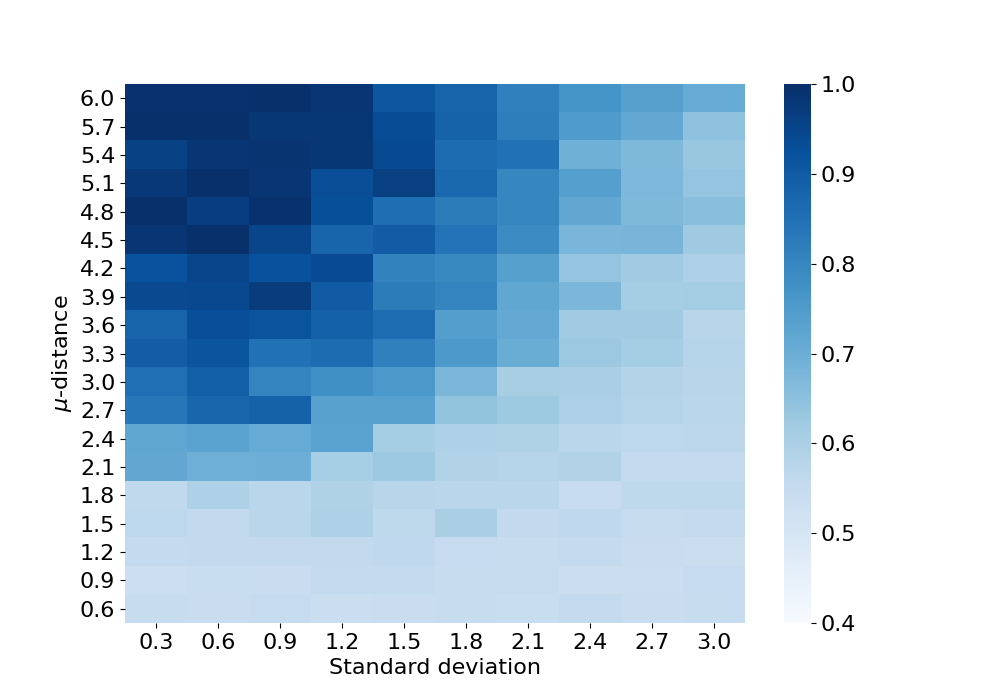}}
\subfloat[]{\includegraphics[width = 9cm]{HC_data_comp_N=100_n_comp=2_total_q=7_noise_q=1_sc=7.png}}}\\
\makebox[0pt]{
\subfloat[]{\includegraphics[width = 9cm]{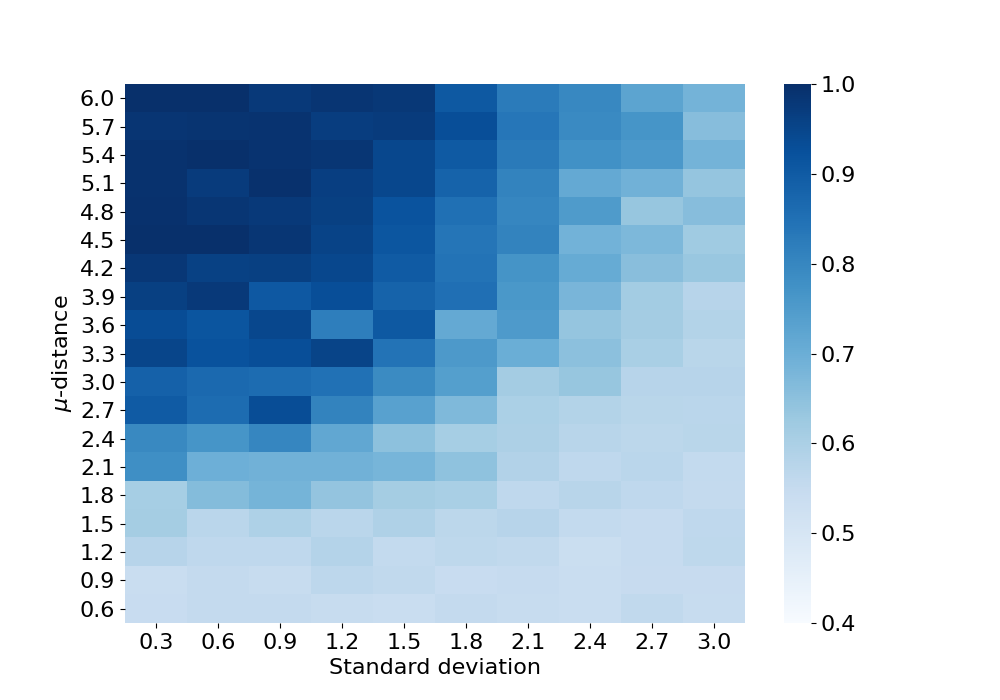}}
\subfloat[]{\includegraphics[width = 9cm]{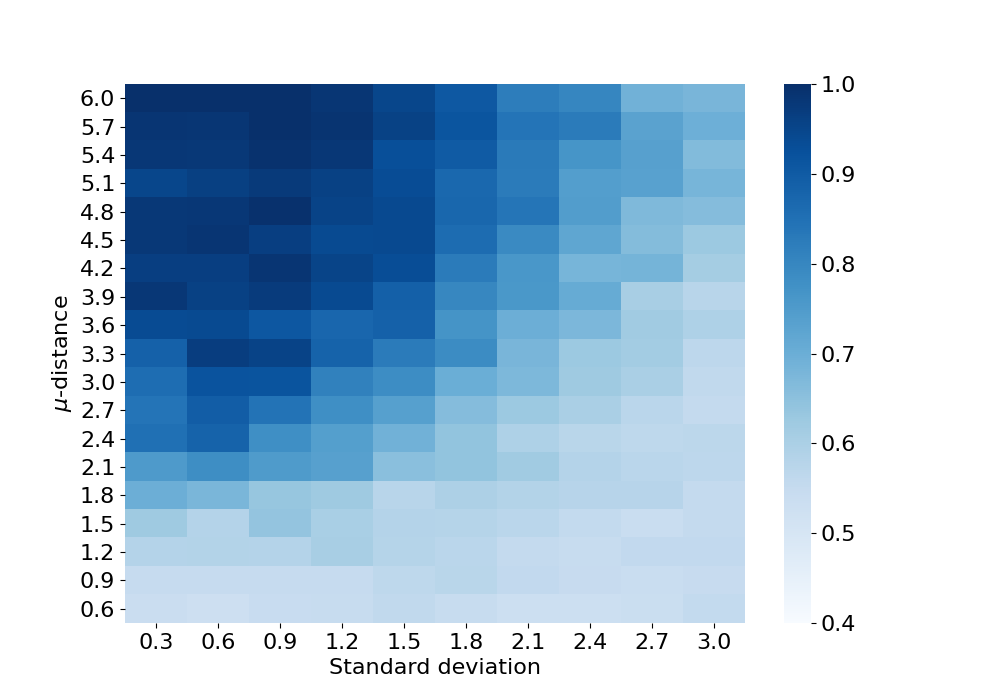}}}

\caption{Heatmaps for the mean correct allocation of hierarchical clustering for synthetic data sets, based on 30 runs per parameter constellation. The four heatmaps only differ in the number of integrated items: a) 6, b) 7, c) 8, d) 9.}
\label{fig:heatmaps_synth_HC}
\end{figure*}

\begin{figure*}
\centering
\makebox[0pt]{
\subfloat[]{\includegraphics[width = 9cm]{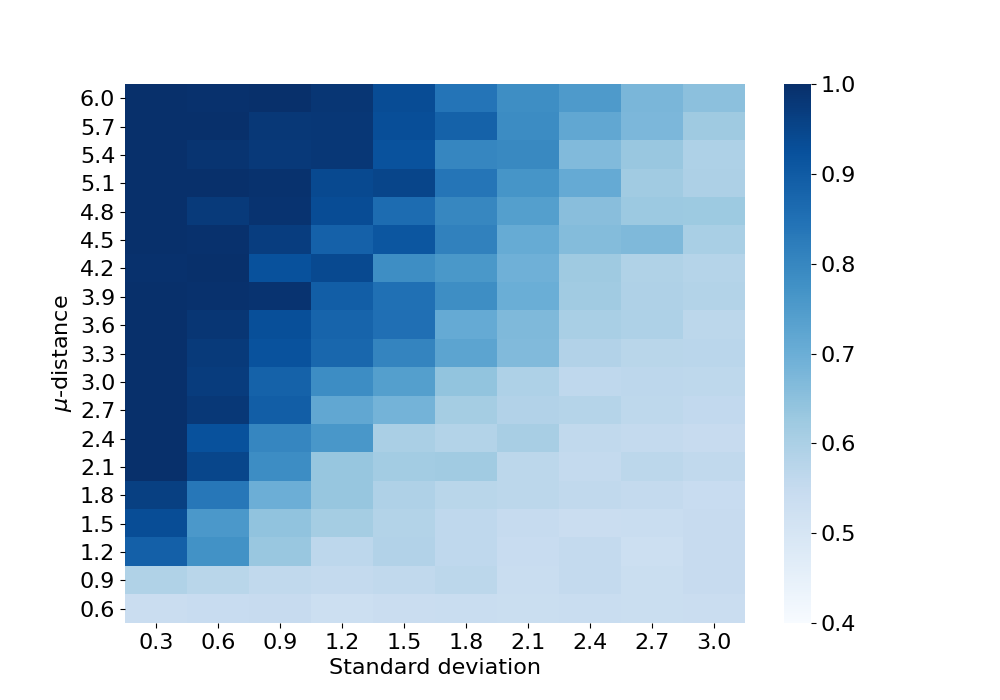}}
\subfloat[]{\includegraphics[width = 9cm]{sbm_data_comp_N=100_n_comp=2_total_q=7_noise_q=1_sc=7.png}}}\\
\makebox[0pt]{
\subfloat[]{\includegraphics[width = 9cm]{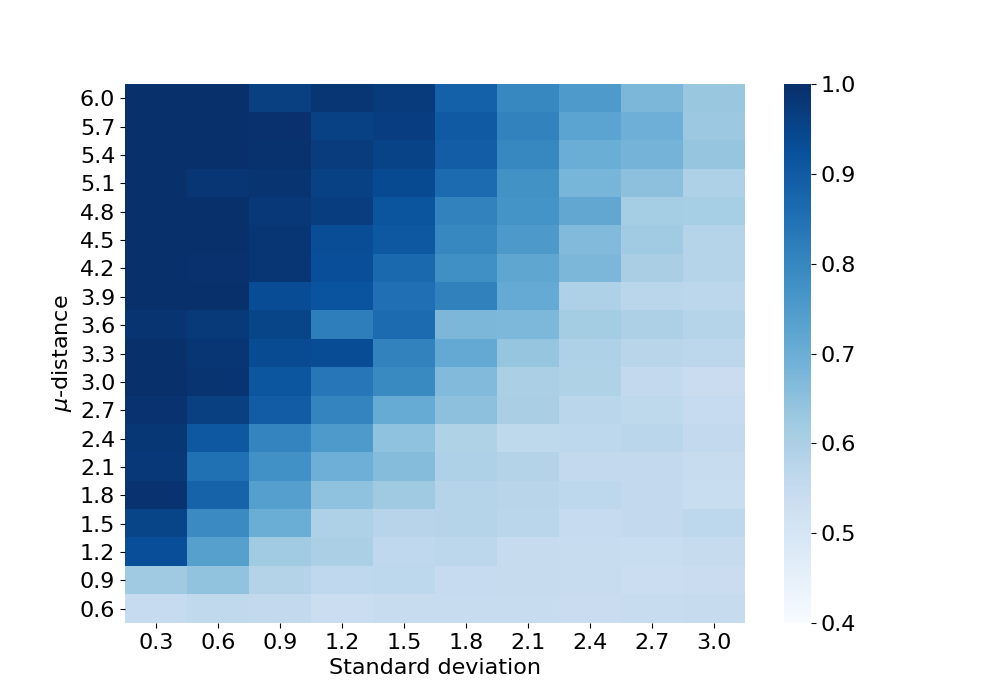}}
\subfloat[]{\includegraphics[width = 9cm]{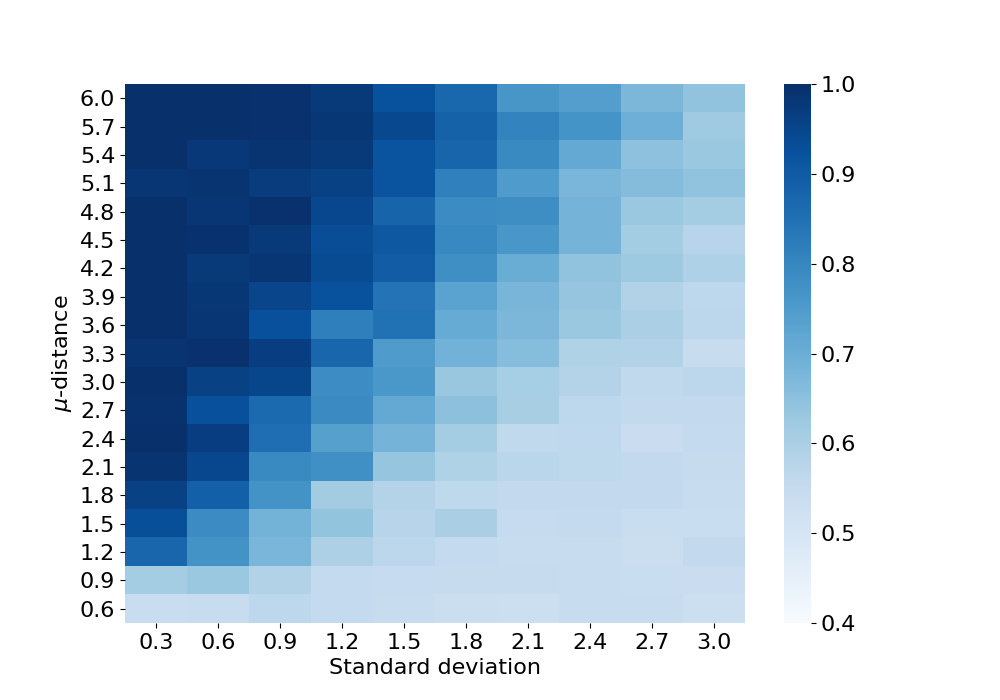}}}

\caption{Heatmaps for the mean correct allocation of stochastic block model for synthetic data sets, based on 30 runs per parameter constellation. The four heatmaps only differ in the number of integrated items: a) 6, b) 7, c) 8, d) 9.}
\label{fig:heatmaps_synth_SBM}
\end{figure*}

In order to examine the performance of the three community detection algorithms, we introduced the synthetic data set construction process. This allows us to to explore the limits of each community detection algorithm. We simulate a large number data sets and run the algorithms on them.\\ 
The synthetic data sets are constructed on artificial results from 100 individuals, with answers to 6, 7, 8 or 9 questions on a scale from 1-7. The questions are ranked in 4 different categories of influence, determined by an increasing mean. The community structure is an equal division into two groups of 50. The \textit{$\mu$-distance} ranges from 0.6 to 6.0 with a step size of 0.3 (y-axis). The corresponding \textit{standard deviation} ranges from 0.3 to 3.0 (x-axis).\\
The heatmaps in Figure~\ref{fig:heatmaps_synth_GN}, \ref{fig:heatmaps_synth_HC} and \ref{fig:heatmaps_synth_SBM} indicate the mean percentage of correctly allocated individuals by the community detection algorithms. Each square of the heatmap represents a mean of 30 simulation runs. For example, the value of 1.0 reports that in 30 simulation runs the algorithm allocated all individuals to the correct community.\\
The results for the Girvan-Newman show the best performance for relatively high \textit{$\mu$-distance} and a low standard deviation. With an increasing standard deviation and a decreasing \textit{$\mu$-distance}, the algorithm is not capable of allocating correctly. The values around 0.5 means that for example a random allocation algorithm would perform likewise. Adding additional items (questions) that are informative to the community structure with-in the simulated data provides slight improvements when trying to recover the communities via the community detection methods (see Figure~\ref{fig:heatmaps_synth_GN}(a-d), dark regions). 
Similar behaviour and results can be seen in the heatmaps for the hierarchical clustering and the stochastic block model (see Figure~\ref{fig:heatmaps_synth_HC} and \ref{fig:heatmaps_synth_SBM}). However, it is striking that the hierarchical clustering method shows a lack of competitiveness, for a low \textit{$\mu$-distance} (between 0.3 and 3.3) and a standard deviation of 0.3 or 0.6. For that parameter group, the results of the Girvan-Newman algorithm and the stochastic block are more convincing.\\

A subsequent step to the analysis of synthetic data and the determination of communities is the evaluation of the questions and their influence on the community structure. 

Within the synthetic data set, we determine the importance of the questions by their overlap of the answer distributions of distinct communities. A higher overlap  means less information concerning the community structure. The question selection method is therefore able to rank the questions by their influence.\\

\begin{figure*}     
\centering
        \makebox[0pt]{
    \subfloat[Girvan-Newman algorithm]{
    \includegraphics[width=\textwidth]{barcharts_Girvan-Newman.png}}}\\
        \makebox[0pt]{
    \subfloat[Hierarchical clustering]{
    \includegraphics[width=\textwidth]{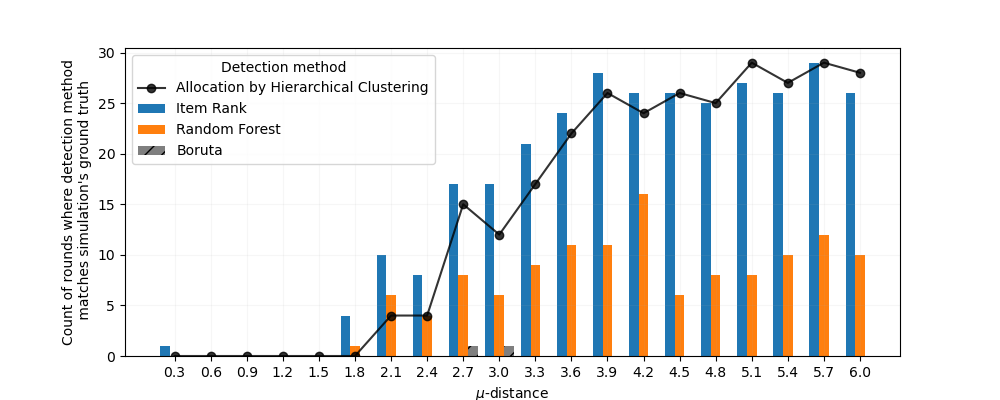}}}\\
        \makebox[0pt]{
    \subfloat[Stochastic block model]{
    \includegraphics[width=\textwidth]{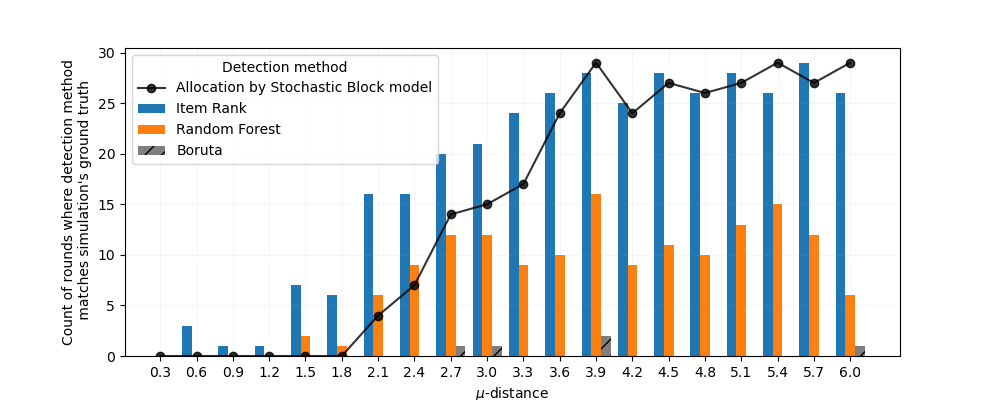}}}
    \caption{Relation between feature selection and community detection algorithms. Each bar represents the simulation results of 30 simulations with the same parameters, the $\mu$-distance is the variable displayed on the x-axis. The lowest standard deviation is 0.7, but it increases for less important questions. The results are based on the communities from: a)~Girvan-Newman algorithm, b) Hierarchical clustering and c) Stochastic block model. The bars show the performance of the questions selection method, Random forest classifier and Boruta (with Random forest classifier). The bar reflects the correct ranking of the questions for the 30 per $\mu$-distance.}
    \label{fig:Compare_feature_selection_HC_GN_SBM}
\end{figure*}

Figure~\ref{fig:Compare_feature_selection_HC_GN_SBM} shows a detail section of the heatmaps. The construction of the synthetic data sets is the same as in the heatmaps, solely the standard deviation is fixed to 0.7. The figure is separated into the results from the three community detection algorithms. Moreover, the ranking results of the question selection method, the Random forest method and the Boruta package are shown. The bars report the number of successful rankings of all 7 questions for 30 runs.  The curve points out the frequency in which the community detection algorithm allocates all individuals of a simulation run correctly. The maximal possible count is 30 for each \textit{$\mu$-distance}.\\
Over all, it is notable that the curves of all three community detection algorithms represent the same dynamics, drawing parallels to the results of the heatmaps. Additionally, the ranking of the questions is related to the performance of the algorithm as it is based on their community allocation. 
For large \textit{$\mu$-distance} (3.6-6), the question selection method classifies in more than 25 cases the ranking of the 7 questions correctly. The Random forest method generally does not exceeds 15. In the cases where the community detection does not work, 1.8 and below, the question selection method and Random forest hardly works. The results for ranking the questions in the case of the Boruta method show that it is not working. It has to be mentioned that the Boruta algorithm focuses on the determination of important and unimportant features or items, and not on the correct ranking of the questions. Nevertheless, the question selection method is able to uncover a high amount of information about the influence of each item.\\
All in all, the question selection method performs very well, which may then justify the long time of execution. However, the Random forest method and the Boruta package are many times faster and therefore applicable on a much larger set of features. 

\section{Results for the ANES data set from 2012 and 2016} \label{appendix:resultsANES}
\begin{figure*} 
    \centering
    \makebox[0pt]{
    \subfloat[Girvan-Newman partition]{
    \includegraphics[width=0.66\textwidth]{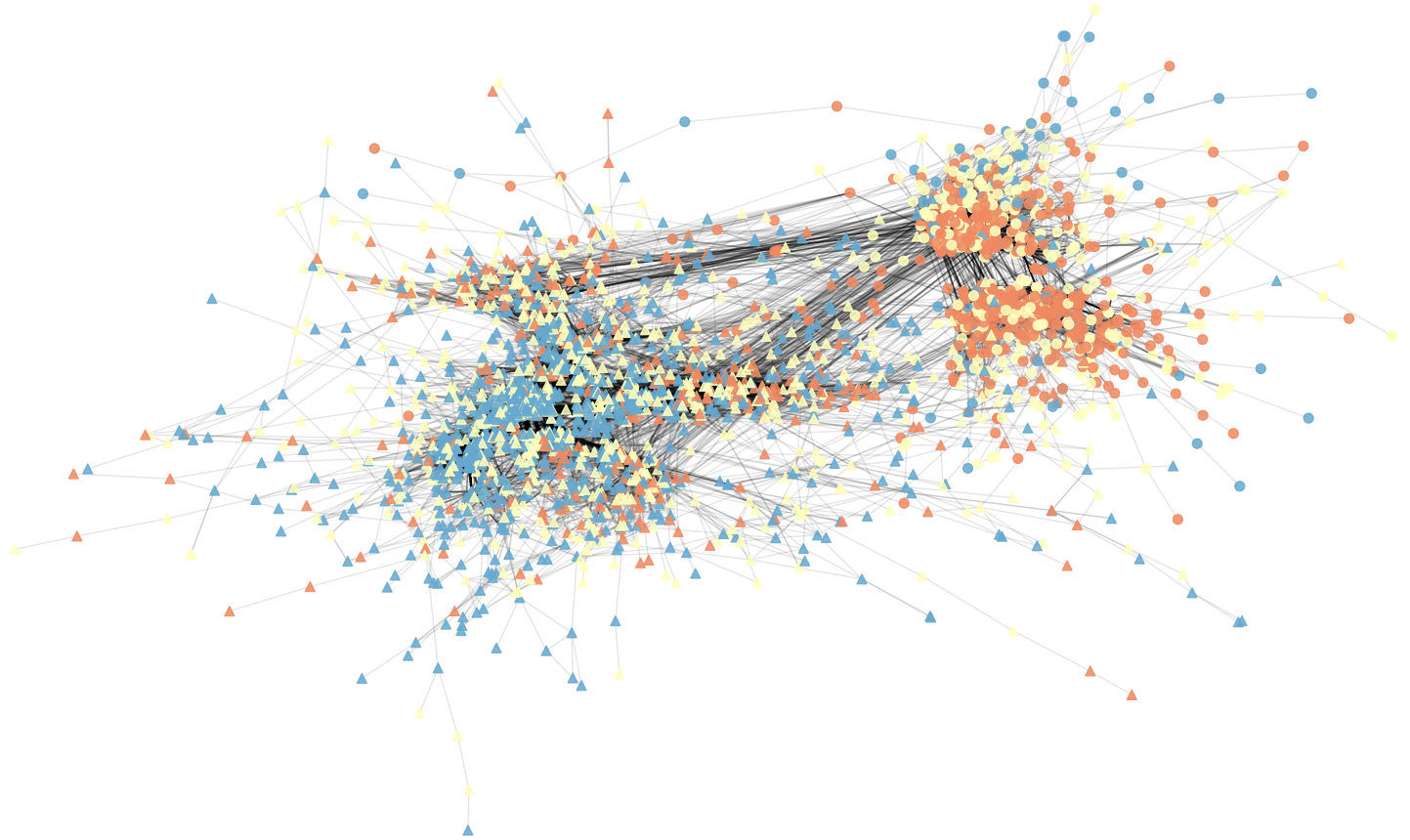}}}\\
        \makebox[0pt]{
    \subfloat[Hierarchical clustering partition]{
    \includegraphics[width=0.66\textwidth]{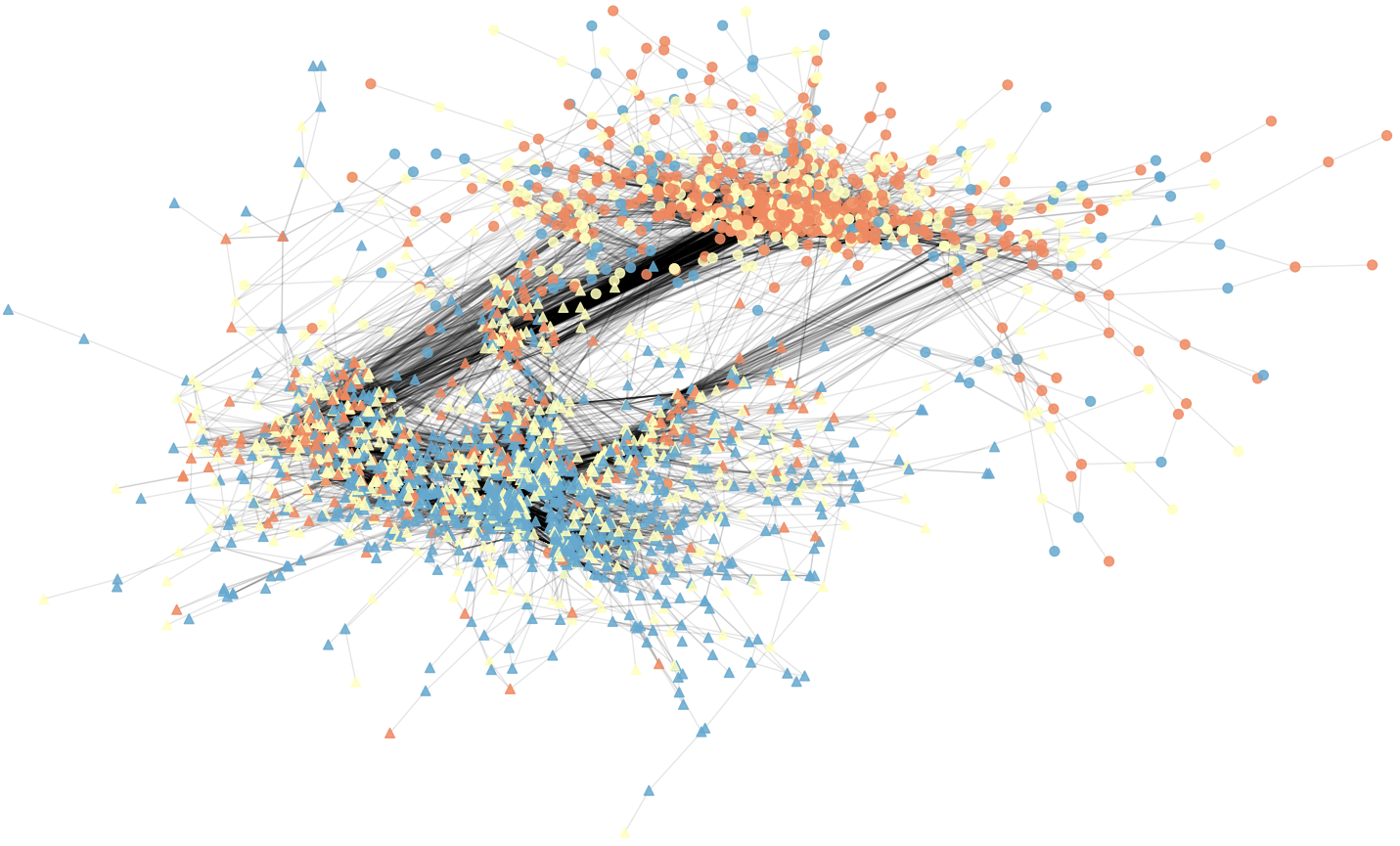}}}\\
        \makebox[0pt]{
    \subfloat[Stochastic block model partition]{
    \includegraphics[width=0.66\textwidth]{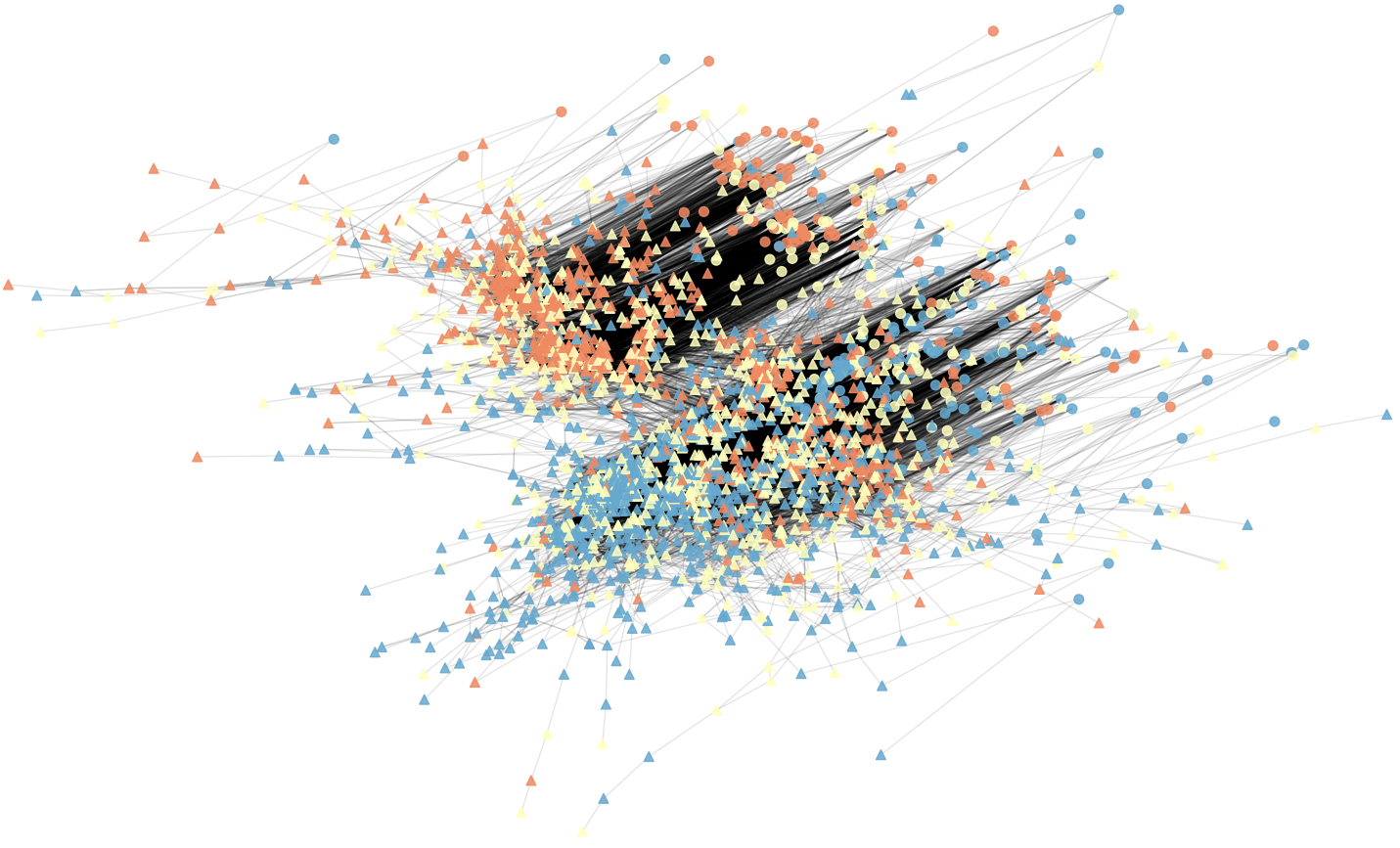}}}
  \caption{American National Election Study data 2012, constructed similarity network with 2 communities, detected by: a) Girvan-Newman algorithm, b) Hierarchical clustering, c) Stochastic block model. The position and shape of the nodes is used to distinguish between the communities. The colour of the nodes represents their party affiliation: republican (red), democrat (blue) and unknown (yellow).}
    \label{fig:network_ANES2012_SI}
\end{figure*}

\begin{figure*} 
    \centering
    \makebox[0pt]{
    \subfloat[Girvan-Newman partition]{
    \includegraphics[width=0.66\textwidth]{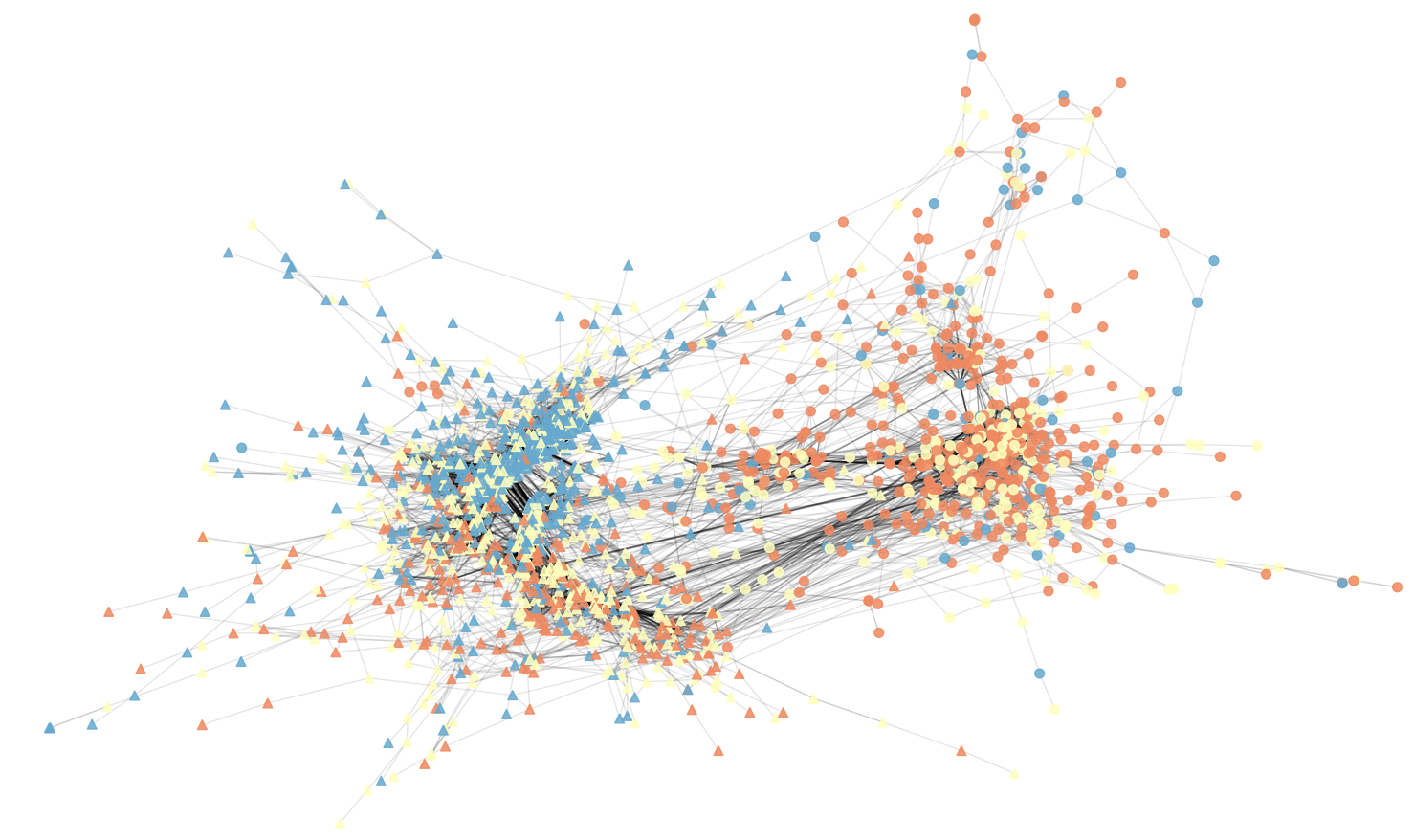}}}\\
        \makebox[0pt]{
    \subfloat[Hierarchical clustering partition]{
    \includegraphics[width=0.66\textwidth]{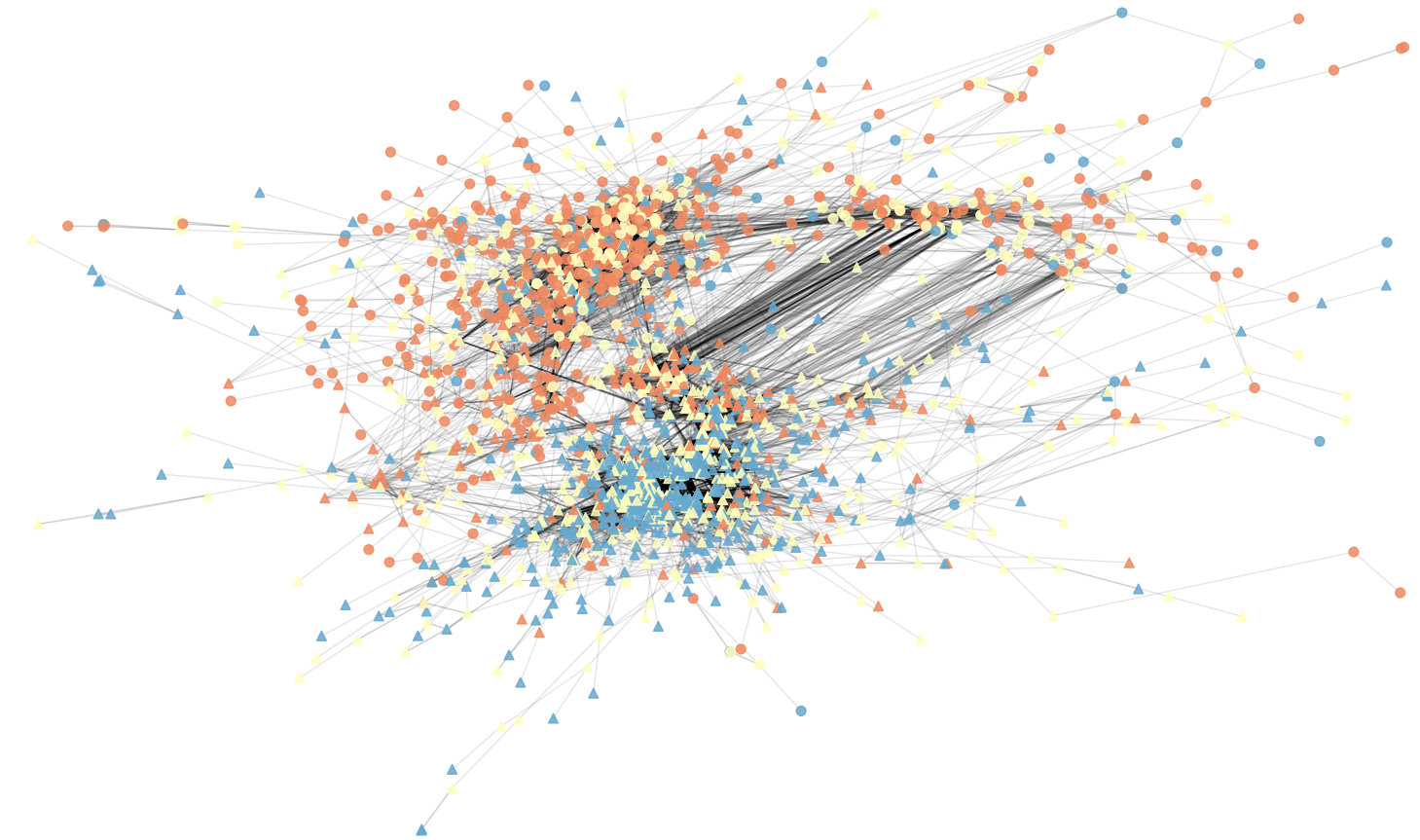}}}\\
        \makebox[0pt]{
    \subfloat[Stochastic block model partition]{
    \includegraphics[width=0.66\textwidth]{SIanes2016_18_02_2021_GN_graph_partition.png}}}
    \caption{American National Election Survey data 2016, constructed similarity network with 2 communities, detected by: a) Girvan-Newman algorithm, b) Hierarchical clustering, c) Stochastic block model. The position and shape of the nodes is used to distinguish between the communities. The colour of the nodes represents their party affiliation: republican (red), democrat (blue) and unknown (yellow).}
    \label{fig:network_ANES2016_SI}
\end{figure*}

The analysis of the ANES data set from 2012 and 2016 was run for the Girvan-Newman algorithm, the hierarchical clustering and the stochastic block model. Only the networks for the Girvan-Newman algorithm were displayed in the main-section. In order to provide the reader with additional information and to be able to compare the network division of the three community detection algorithms, the networks are shown here (see Figure~\ref{fig:network_ANES2012_SI} \& \ref{fig:network_ANES2016_SI}).

\paragraph{Reduced results for ANES data set 2016} \label{appendix:resultsANESminusimmigration}
The analysis of the ANES data set from 2012 and 2016 was run for the Girvan-Newman algorithm, the hierarchical clustering and the stochastic block model. Only the networks for the Girvan-Newman algorithm were displayed in the main-section. In order to provide the reader with additional information and to be able to compare the network division of the three community detection algorithms, the networks are shown here.

\begin{figure*} 
    \centering
    \makebox[0pt]{
    \subfloat[Girvan-Newman partition]{
    \includegraphics[width=0.66\textwidth]{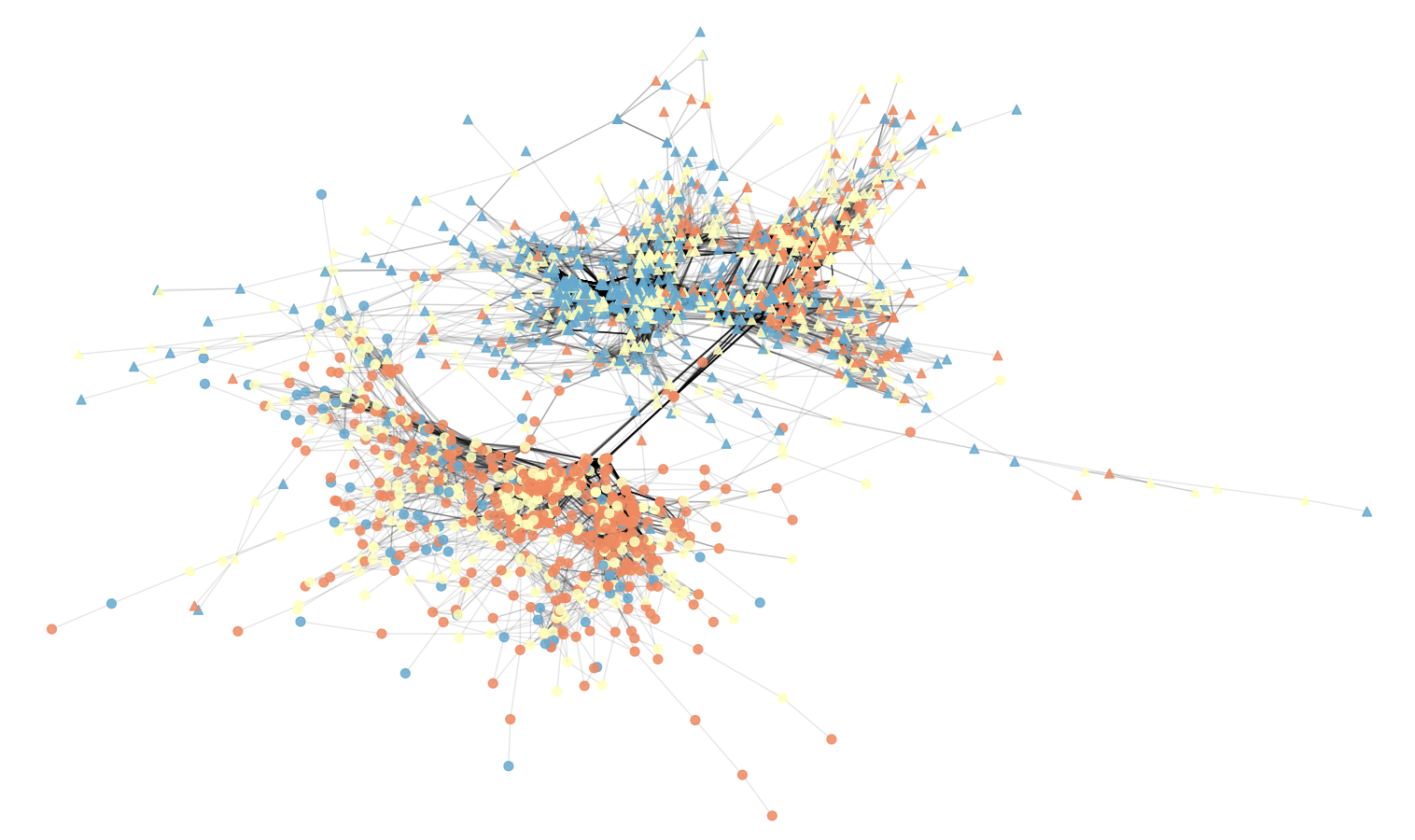}}}\\
        \makebox[0pt]{
    \subfloat[Hierarchical clustering partition]{
    \includegraphics[width=0.66\textwidth]{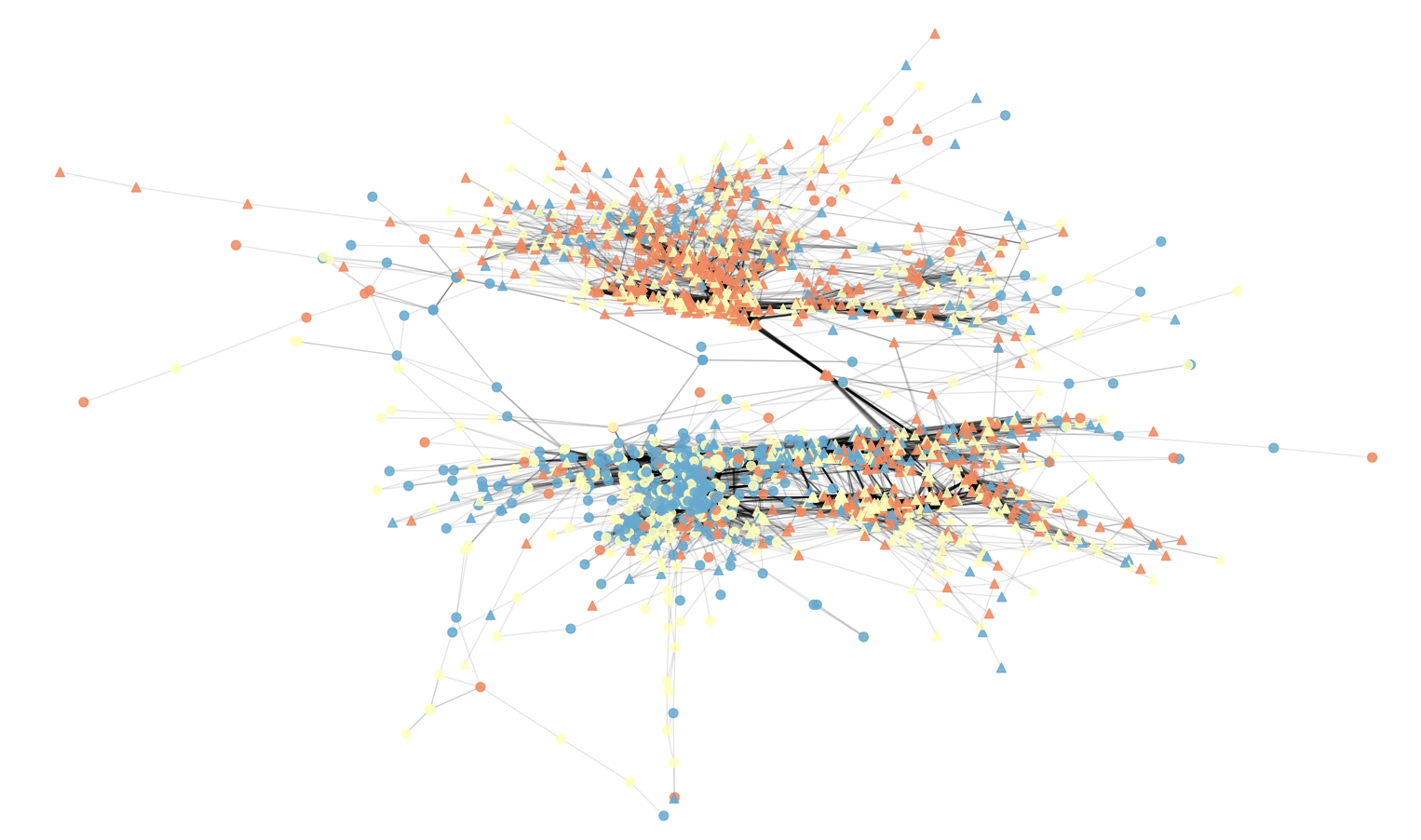}}}\\
        \makebox[0pt]{
    \subfloat[Stochastic block model partition]{
    \includegraphics[width=0.66\textwidth]{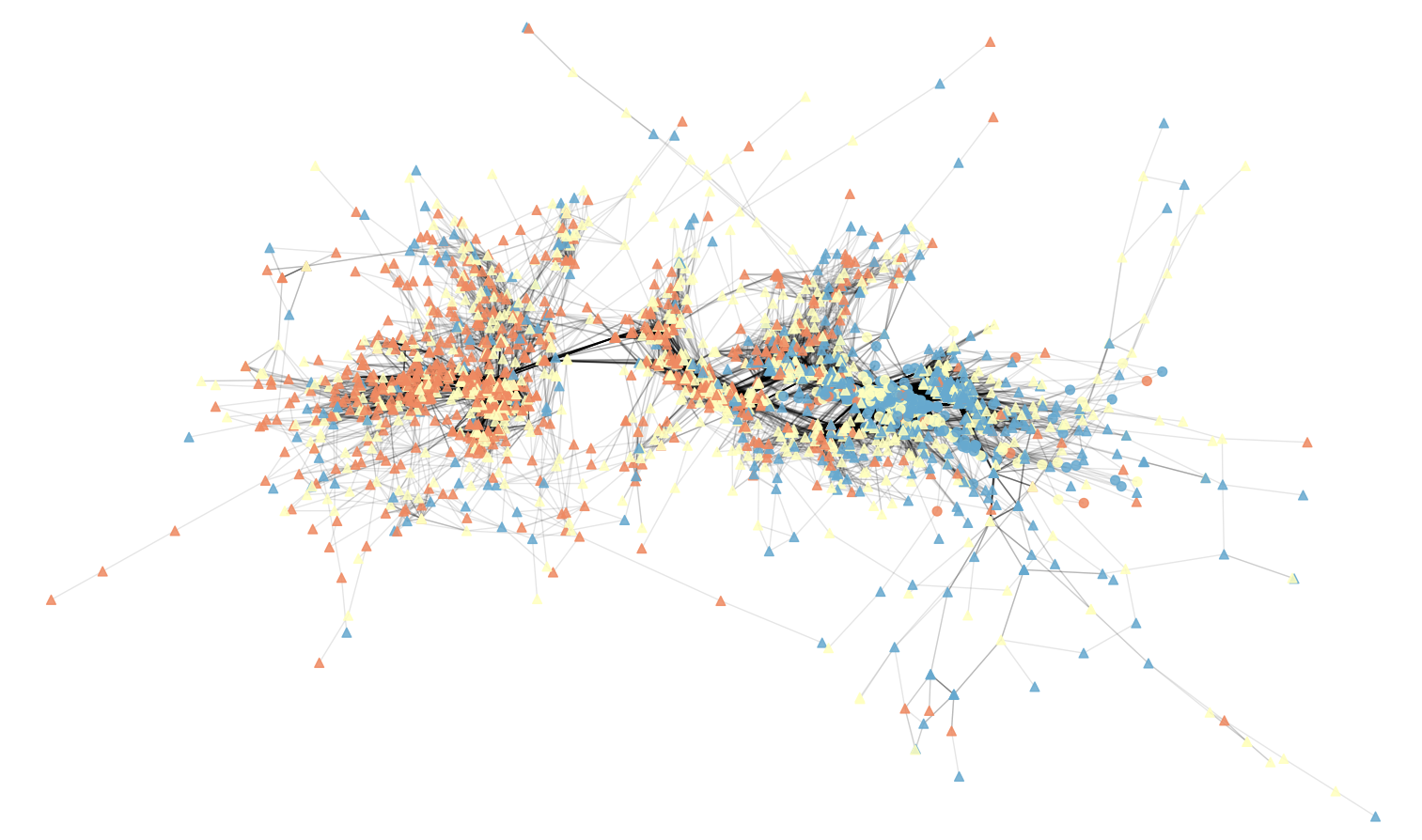}}}
    \caption{American National Election Survey data 2016, constructed similarity network with 2 communities, detected by: a) Girvan-Newman algorithm, b) Hierarchical clustering, c) Stochastic block model. We reduced the number of variables from 8 to 7. The shape of the nodes is used to distinguish between the communities. The colour of the nodes represents their party affiliation: republican (red), democrat (blue) and unknown (yellow).}
    \label{fig:network_ANES2016_SI_reduced}
\end{figure*}

To confirm the communities from the ANES data set 2016 in Figure~\ref{fig:ANES2016_GN} and to test its robustness, we reduced the used variables to construct the network and the communities. We rerun the analysis for the same set of variables but without the variable \textit{Immigration}. We selected the variable \textit{Immigration} to be eliminated because according to the item rank method it has the least impact on the community structure.\\
Figure~\ref{fig:network_ANES2016_SI_reduced} displays two communities with similar sizes and similar partisan sorting like in Figure~\ref{fig:ANES2016_GN}. The sizes of the communities of the Girvan-Newman algorithm are 1879 and 966. The reduction from eight to seven variables did not change the overall structure of the network.


\section{Additional results for Singapore}
From the Wellcome Global Monitor, we used all items which express positions to several institutions and vaccines. Here we reapply the method and only use the available data without the items on vaccines. We show that it is possible to receive similar communities without the vaccine questions as they are not listed at the top of item ranking.\\
The reduction of the data leads to more individuals who answered all items. Therefore, the size of the constructed network increased to 531. Nonetheless, the overall structure remains very similar to the results with the vaccine items. The community split up is 386 and 145 individuals and they are only connected by 5 cross-cutting links. The threshold is 9 and the total amount of links is 7136. The method detects two opinion-based groups that are connected within the groups and have nearly no connection between groups.\\\\
The item ranking provides us with the information about the influence on the community structure. For an additional analysis, we investigated how many of the most relevant items do we need to get a similar network structure and communities.
We successively added an item until we received a similar network structure. We integrated the first 9 out of 13 items to get to a similar network structure. With a threshold of 8.2, the method constructed a network with a size of 597, with two communities (492, 105) and a total number of 8270 links where only 2 of them were cross-cutting links.

\section{Threshold and the giant component}

\begin{figure*}
\includegraphics[width=0.48\textwidth]{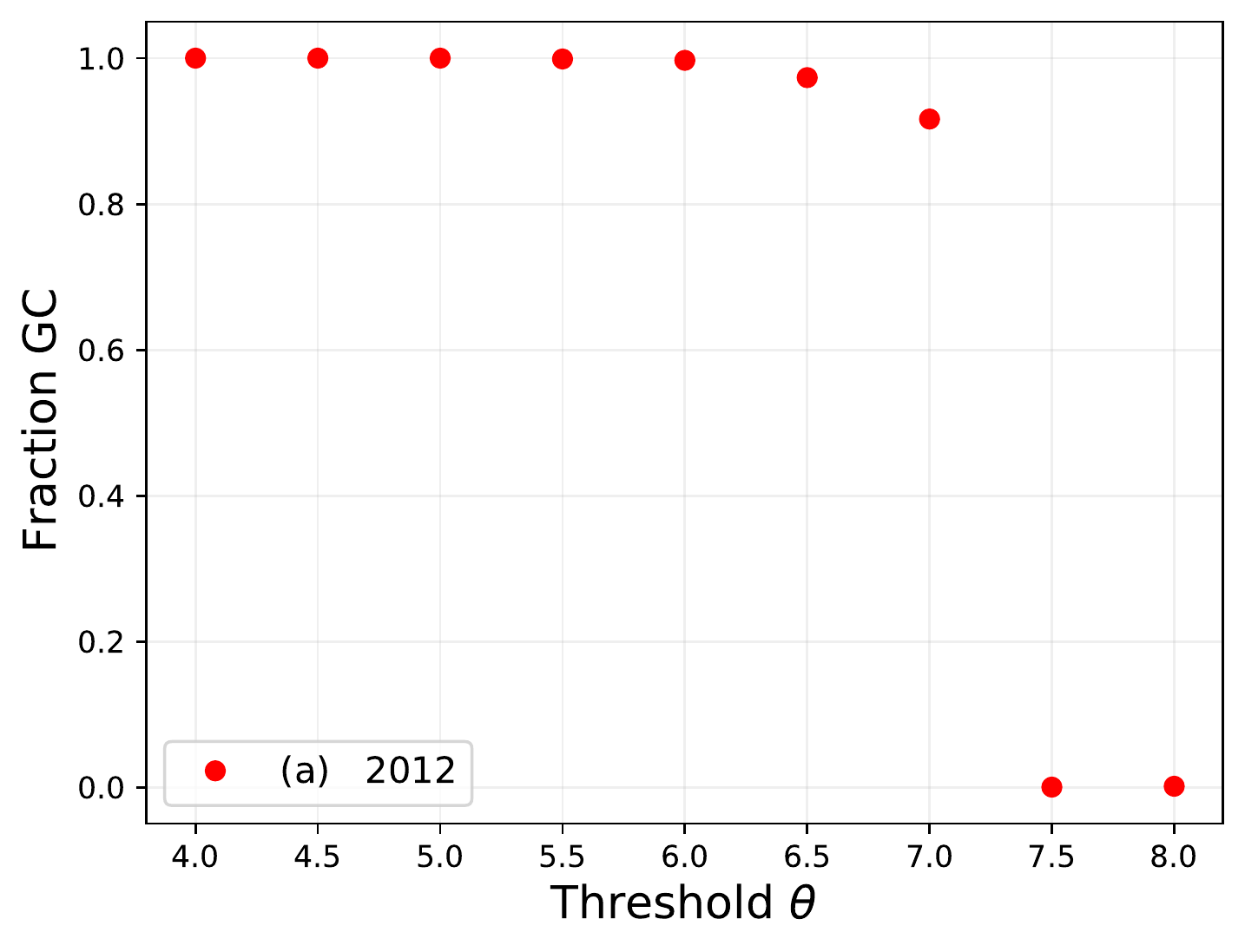} 
\includegraphics[width=0.48\textwidth]{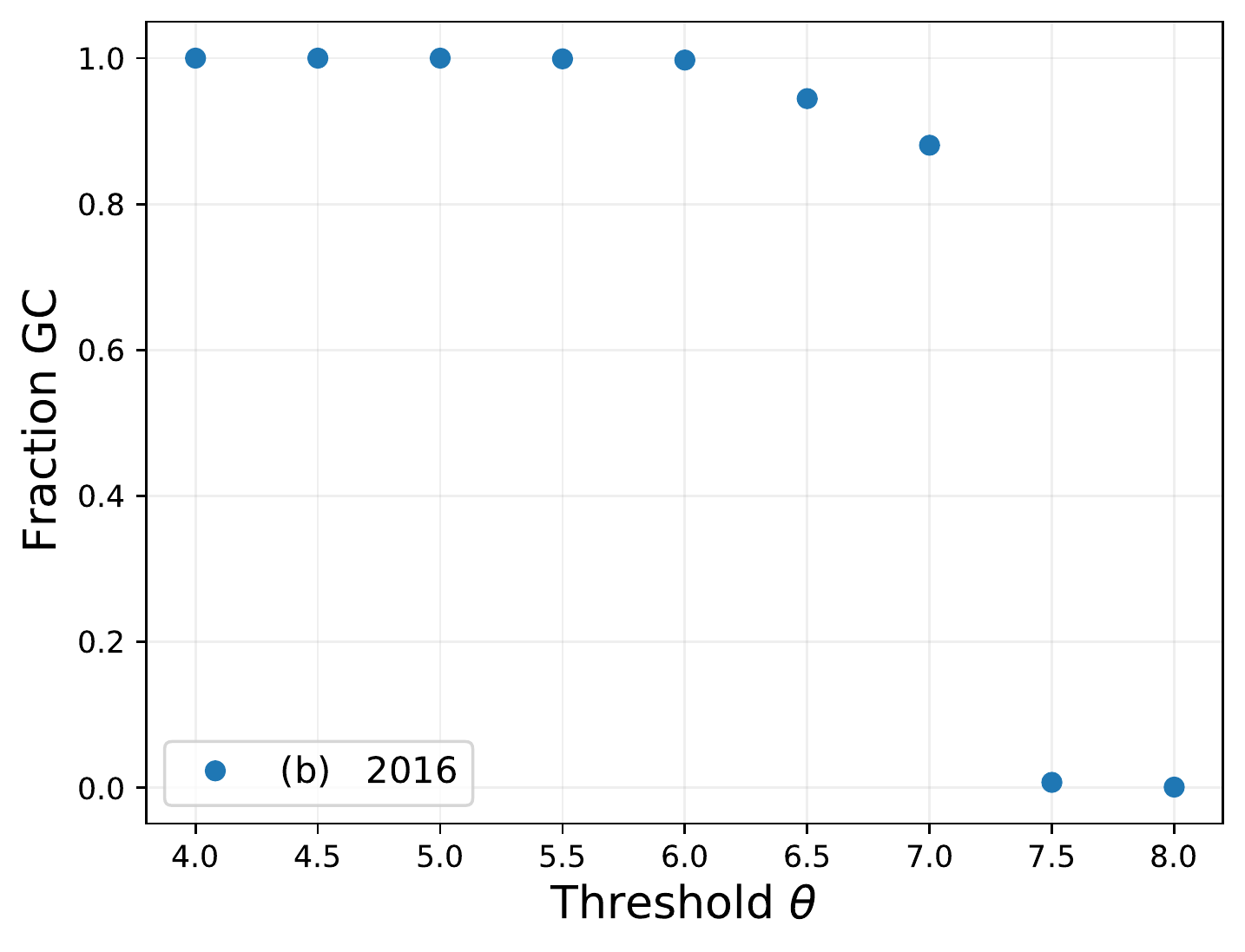} 
\caption{The fraction of the giant component versus the agreement threshold $\theta$ increasing in intervals of 0.5 for the ANES data in (a) 2012 and (b) 2016. As can be seen, there is a sudden jump from $\theta=7.5$ to $\theta=7$.}
    \label{fig:Threshold}
\end{figure*}

To decide the threshold for agreement between participants $\theta$ we plot the fraction of participants in the giant component against $\theta$ in Figure~\ref{fig:Threshold} for the ANES data in 2012 and 2016. In each case when $\theta=7$ there is a jump in the number of nodes in the giant component and then it slowly increases to include all participants. Hence, we use this value of $\theta$ as the network is not too dense yet includes the majority of the nodes. Further decreasing $\theta$ vastly increases the number of links. We find similar communities in this case but it is less efficient due to the high density of links.

\end{document}